# Current-driven dynamics of chiral ferromagnetic domain walls

Satoru Emori[1], Uwe Bauer[1], Sung-Min Ahn[1], Eduardo Martinez[2], and Geoffrey S. D. Beach[1*]

[1]Department of Materials Science and Engineering, Massachusetts Institute of Technology,

Cambridge, Massachusetts 02139, USA

[2]Dpto. Física Aplicada. Universidad de Salamanca,

Plaza de los Caidos s/n E-38008, Salamanca, Spain

**In most ferromagnets the magnetization rotates from one domain to the next with no preferred handedness. However, broken inversion symmetry can lift the chiral degeneracy, leading to topologically-rich spin textures such as spin-spirals[1,2] and skyrmions[3–5] via the Dzyaloshinskii-Moriya interaction (DMI)[6]. Here we show that in ultrathin metallic ferromagnets sandwiched between a heavy metal and an oxide, the DMI stabilizes chiral domain walls (DWs)[2,7] whose spin texture enables extremely efficient current-driven motion[8–11]. We show that spin torque from the spin Hall effect[12–15] drives DWs in opposite directions in Pt/CoFe/MgO and Ta/CoFe/MgO, which can be explained only if the DWs assume a Néel configuration[7,16] with left-handed chirality. We directly confirm the DW chirality and rigidity by examining current-driven DW dynamics with magnetic fields applied perpendicular and parallel to the spin spiral. This work resolves the origin of controversial experimental results[10,17,18] and highlights a new path towards interfacial design of spintronic devices.**

*e-mail: gbeach@mit.edu

Current-controlled DW displacement underpins the operation of an emerging class of spintronic memory[19] and logic[20,21] devices. In out-of-plane magnetized ferromagnets sandwiched between an oxide and a heavy metal, current-induced DW motion is anomalously efficient.[8–11] This observation has been widely attributed to a Rashba effective field[17,22,23] that stabilizes Bloch DWs against deformation, permitting high-speed motion[10] via conventional spin-transfer torque (STT)[24]. However, current-induced DW motion is absent in symmetric Pt/Co/Pt[8,9,11,25] stacks, and semiclassical transport calculations[25] suggest the spin-polarized current in the ultrathin (< 1 nm) Co is vanishingly small. Moreover, DWs in Pt/Co/oxide move against electron flow[8,10,11], contrary to the action of STT[24]. Together, these results suggest that conventional STT contributes negligibly to DW dynamics in these ultrathin structures and that interfacial phenomena[26,27] are instead responsible.

The Rashba field lacks the correct symmetry to drive DWs directly[16,26,27], and the spin Hall effect (SHE) in the adjacent heavy metal has emerged as a possible alternative mechanism[12–16,27]. SHE-driven spin accumulation at the heavy-metal/ferromagnet interface generates a Slonczeswki-like torque[16,26,27] strong enough to switch uniformly-magnetized films[12–15,18]. However, the Bloch DWs expected in typical nanowire geometries[8–11,28] have their plane oriented perpendicular to the nanowire axis, in which case the Slonczewski-like torque vanishes[16]. This behavior was recently confirmed in asymmetric Pt/Co/Pt stacks in which the SHE-induced torques from the Pt layers did not cancel completely[15]. In that case, current-assisted DW depinning was observed when an applied field rotated the DW plane towards the current axis, but up-down and down-up DWs were driven in opposite directions and the current had no effect in the absence of the bias field. The SHE alone is therefore incapable of uniformly driving trains of DWs in devices, and is insufficient to explain the high spin-torque efficiencies and DW velocities observed in Pt/Co/oxide[8–11] without applied fields.

Here we characterize current-induced torques and DW dynamics in out-of-plane magnetized Pt/CoFe/MgO and Ta/CoFe/MgO stacks that are nominally identical except for the heavy-metal underlayers, whose spin Hall angles are large and of opposite sign[12–14]. By considering the symmetry of the measured current-induced torque along with the DW dynamics driven by this torque, we uniquely identify the DW configuration as Néel with a fixed chirality. Magnetostatics alone makes this configuration unstable and does not favor one chirality over the other, but the DMI has been theoretically shown to promote chiral Néel DWs[2,7]. By applying in-plane magnetic fields, we verify that the DW magnetization aligns rigidly along the nanowire axis, and that the DW spin spiral exhibits a global chirality common to both Pt/CoFe/MgO and Ta/CoFe/MgO. Current-driven DW motion in heavy-metal/ferromagnet/oxide structures is naturally explained by the combination of the SHE, which produces the sole current-induced torque, and the DMI, which stabilizes chiral DWs whose symmetry permits uniform motion with very high efficiency.

DW motion was characterized in 500-nm wide, 40-μm long nanowires overlaid with an orthogonal DW nucleation line and lateral contacts for current injection (Fig. 1a). We first examine the effect of current on the threshold field $H_\text{prop}$ for DW propagation through the defect landscape. Measurements were performed as in Ref. 11, by first nucleating a reversed domain with the Oersted field from a current pulse through the nucleation line and then sweeping an out-of-plane field $H_z$ to drive the DW along the nanowire. DW motion was detected via the polar magneto-optical Kerr effect, with a ~3 μm laser spot positioned at the midpoint of the nanowire. Comparing Figs. 1d,e, $H_\text{prop}$ varies linearly with electron current density $j_e$, but DW propagation is hindered in the electron flow direction in Pt/CoFe/MgO and assisted along electron flow in Ta/CoFe/MgO. This remarkable difference, produced simply by changing the nonmagnetic metal in contact with the ferromagnet, was independent of the sense of magnetization (up-down or down-up) across the DW. The magnitude of the spin-torque efficiency, taken as the slope of $H_{prop}$ versus $j_e$, was 120 Oe/$10^{11}$ A m$^{-2}$ for Pt/CoFe/MgO and 170 Oe/$10^{11}$ A m$^{-2}$ for Ta/CoFe/MgO. These large efficiencies are comparable to those reported for Pt/Co/AlOx[9,10] and Pt/Co/GdOx[11], suggesting that a universal mechanism governs current-driven DW motion in heavy-metal/ferromagnet/oxide.

In Figs. 1f,g, we directly compare field-driven and current-driven DW velocities, measured using a time-of-flight technique[11]. Again, DWs moved against electron flow in Pt/CoFe/MgO (Fig. 1f) and along electron flow in Ta/CoFe/MgO (Fig. 1g). The maximum field was limited by random domain nucleation, and the exponential dependence of velocity on $H_z$ and $j_e$ indicates thermally activated motion[11,29]. The field-driven and current-driven velocities exhibit the same dynamical scaling across three decades in velocity when $j_e$ is scaled by a constant (110 Oe/$10^{11}$ A m$^{-2}$ for Pt/CoFe/MgO and 160 Oe/$10^{11}$ A m$^{-2}$ for Ta/CoFe/MgO). These field-to-current ratios closely match those extracted from Figs. 1d,e. We therefore conclude that the effect of current on DW motion is phenomenologically equivalent to an out-of-plane field[9,11], which reveals the symmetry of the current-induced torque as discussed later in this Letter.

In addition to robust DW motion, current enables switching between uniformly magnetized "up" and "down" states with the assistance of a constant in-plane magnetic field[12,15,18]. This switching phenomenon was demonstrated in 1200-nm wide Hall crosses (Fig. 2a). A sequence of 250-ms long current pulses with increasing (or decreasing) amplitude was injected along the *x*-axis, and in between each pulse the out-of-plane magnetization component $M_z$ was measured from the anomalous Hall voltage using a low-amplitude (~$10^9$ A m$^{-2}$) AC sense current and a lock-in amplifier. Figs. 2d,e plot $M_z$ versus $j_e$, under a constant applied longitudinal field $H_L$. This field tilted the magnetization away from the z-axis by $\approx 5°$ in Pt/CoFe/MgO at 500 Oe and $\approx 3°$ in Ta/CoFe/MgO at 100 Oe, but did not bias $M_z$ up or down, as evidenced by the nearly symmetric switching profile (Figs. 2d,e). With sufficiently large $H_L$ and $j_e$ in the +x direction, the up magnetized state was favored in Pt/CoFe/MgO (Fig. 2d, solid line), whereas the down state was favored in Ta/CoFe/MgO (Fig. 2e, solid line). When the direction of $H_L$ or $j_e$ was reversed, the preferred magnetization direction was also reversed (Figs. 2d, e, dotted lines).

This switching behavior implies that $j_e$ generates an effective field $\vec{H}_{SL}$ associated with a Slonczewski-like torque[12,13,15,18], given by $\vec{H}_{SL} = H_{SL}^0 (\hat{m} \times (\hat{z} \times \hat{j}_e))$[16]. Here $\hat{m}$, $\hat{z}$, and $\hat{j}_e$ are unit vectors along the magnetization, z-axis, and electron flow, respectively, and $H_{SL}^0$ parameterizes the torque. The SHE in the heavy metal directly generates a Slonczewski-like torque, but the Rashba effect can also yield a torque of this form due to spin-relaxation[18,26,27]. Assuming the SHE is the dominant source, justified experimentally below, $H_{SL}^0$ is related to the spin Hall angle $\theta_{SH}$ in the heavy metal via $H_{SL}^0 = \hbar \theta_{SH} |j_e|/(2|e|M_S t_F)$[16], with $M_S$ the saturation magnetization and $t_F$ the ferromagnet thickness. From the sign of $H_{SL}^0$ extracted from current-induced switching (Figs. 2b,c), $\theta_{SH}$ is positive in Pt and negative in Ta, consistent with Refs. 12 and 13.

We quantified the Slonczewski-like torque by detecting current-induced magnetization tilting using the technique in Ref. 30. AC current exerts a periodic torque on the uniformly magnetized state, causing $M_z$ to vary at the drive frequency, $\omega$. This leads to first and second harmonics in the anomalous Hall voltage, $V_\omega$ and $V_{2\omega}$, from which the magnetization tilting can be determined[22,30]. We measured $V_\omega$ and $V_{2\omega}$ while quasistatically sweeping a longitudinal field $H_L$, and extracted $H_{SL}$ via $H_{SL} = -2(dV_{2\omega}/dH_L)/(d^2V_\omega/dH_L^2)$ (Ref. 30, Supplementary Information). Measurements were performed at several AC amplitudes to extract the scaling of $H_{SL}$ with current. When the magnetization was up and $j_e$ was in the +x direction, $H_{SL}$ pointed along -x in Pt/CoFe/MgO (Figs. 3a,e) and +x in Ta/CoFe/MgO (Figs. 3b,f), in agreement with our analysis of magnetization switching (Figs. 2b, c). The direction of $H_{SL}$ reversed when the magnetization was oriented down. The linear fit in Fig. 3a reveals a large $H_{SL}^0$ in Pt/CoFe/MgO of magnitude 50 Oe per $10^{11}$ A m$^{-2}$, implying $\theta_{SH} = +0.06$ in Pt, which agrees well with Ref. 12. The magnitude of $H_{SL}^0$ in Ta/CoFe/MgO is $\approx 200$ Oe per $10^{11}$ A m$^{-2}$, implying $\theta_{SH} = -0.25$ in Ta, twice as large as in Ref. 13 and closer to the value reported for W[14].

The current-induced effective transverse field $H_{FL}$, often associated with a "field-like" torque from the Rashba effect[16,17,22,23,26,27], was quantified similarly by sweeping an applied transverse field $H_T$: $H_{FL} = -2(dV_{2\omega}/dH_T)/(d^2V_\omega/dH_T^2)$. Unlike $H_{SL}$, the direction of $H_{FL}$ was independent of the magnetization orientation (Figs. 3 c,d). The magnitude of $H_{FL}$ in Pt/CoFe/MgO (Fig. 3c) was $\approx 20$ Oe/$10^{11}$ A m$^{-2}$, two orders of magnitude lower than reported in Refs. 17 and 22, although its directionality was the same as in Pt/Co/AlOx[17,22]. Since current-induced DW motion had a very high efficiency and occurred against the electron flow direction in Pt/CoFe/MgO, the fact that $H_{FL}$ was negligible indicates that the Rashba effect cannot be the source of these features[8–11,26,27]. Furthermore, since any contribution to the Slonczewski-like torque by the Rashba effect[18] enters as a correction proportional to the nonadiabicity parameter $\beta \ll 1^{26,27}$, the fact that $H_{SL}$ is here much larger than $H_{FL}$ implies that the Rashba effect contributes negligibly to the Slonczewski-like torque.

In Ta/CoFe/MgO (Fig. 3d), $H_{FL}$ was by contrast quite large, $\approx 400$ Oe/$10^{11}$ A m$^{-2}$, and its direction was the same as in Ta/CoFeB/MgO[23] and opposite to Pt/CoFe/MgO and Pt/Co/AlOx[17,22], suggesting a strong Rashba field[17,22,23] in this sample. However, as noted in Ref. 12, macrospin modeling shows that a large Slonczewski-like torque can pull the magnetization out of the *x-z* plane, which in the measurements here and elsewhere[17,22,23,30] would have a similar effect as a field-like torque. Considering the weaker perpendicular magnetic anisotropy (see Methods) and the much larger $H_{SL}^0$ in Ta/CoFe/MgO compared to Pt/CoFe/MgO, the Slonczewski-like torque could contribute to the apparently large $H_{FL}$.

As summarized in Figs. 3e,f, the current-induced torques are opposite in Pt/CoFe/MgO and Ta/CoFe/MgO, as are the direction of current-driven DW motion and the sign of the spin Hall angles in Pt and Ta. Here we consider in detail the case of Pt/CoFe/MgO, in which the field-like torque is unambiguously small. One-dimensional (1D) model calculations[29] in Fig.4b (see Methods and Supplementary Information) show that Bloch DWs cannot be driven by the SHE alone, in agreement with prior reports[16,27] and with the symmetry of the Slonczewski-like torque. In the 1D model with $\theta_{SH} > 0$ and with no transverse Rashba field, the addition of conventional STT enables sustained DW motion, but its direction is along electron flow (Fig. 4b). No combination of the SHE and STT reproduces the experimentally-observed DW motion against electron flow (Supplementary Information), and moreover conventional STT is likely absent as argued above. Thus, an alternate mechanism is required whereby the SHE alone can drive DW motion.

Néel DWs have an internal magnetization that would align with the nanowire axis, such that the Slonczewski-like torque would manifest as a *z*-axis field[16] as experimentally observed (Fig. 1). However, the direction of $H_{SL}$ depends of the sense of the DW magnetization, and the direction of DW motion varies accordingly (Supplementary Information). Fig. 4a illustrates Néel DWs with oppositely-directed internal magnetization for up-down and down-up transitions, exhibiting a left-handed chiral texture[2]. Based on the sign of the measured Slonczewski-like torque (Figs. 2 and 3), these chiral DWs move against electron flow in Pt/CoFe/MgO and along electron flow in Ta/CoFe/MgO. Although Bloch DWs are magnetostatically preferred[28], adding the DMI to the 1D model stabilizes such chiral Néel DWs[7] (Methods, Supplementary Information), leading to qualitative behavior in agreement with experiment (Fig. 4b).

Finally, we assess the rigidity and chirality of the Néel DWs in Pt/CoFe/MgO using applied in-plane fields. In Figs. 4c,d we show that the spin-torque efficiency, extracted similarly to Fig. 1d, is insensitive to $H_L$ up to at least 600 Oe, but declines significantly with increasing $|H_T|$. This behavior is opposite to that reported for Bloch DWs in Ref. 15, but is precisely what is expected for DMI-stabilized Néel DWs: $H_L$ is collinear with the DW magnetization and exerts no torque, whereas $H_T$ exerts a torque that cants the DW magnetization away from the *x*-axis and reduces the *z*-axis-oriented $H_{SL}$. That the sense of internal DW magnetization could not be reversed at the experimentally-available maximum $H_L$ of 600 Oe attests to the strength of the DMI in this system.

We also measured the effects of $H_L$ and $H_T$ on the velocity of fast current-driven DWs (Figs. 4e,f), which was reproduced qualitatively by the 1D model with the SHE and DMI (Figs. 4 g,h). $H_L$ modified the velocities of up-down and down-up DWs with opposite slopes (Figs. 4e,g), whereas $H_T$ modified both velocities identically (Figs. 4f,h). The 1D model predicts DW motion reversal under very large $H_L$ coinciding with reversal of the DW sense, and impeded motion for large $H_T$ due to rotation towards a Bloch configuration (see Supplementary Information). Interestingly, the velocity increased with $H_T$ in the direction of the previously reported Rashba field in Pt/Co/AlOx[10,17,22], although here $H_{FL}$ in Pt/CoFe/MgO was vanishingly small. Our experimental and computational results indicate that, even without the Rashba effect, $H_T$ can modify the dynamics of Néel DWs driven by the SHE-induced Slonczewski-like torque.

In summary, we show that current alone drives DWs with high efficiency but in opposite directions in Pt/CoFe/MgO and Ta/CoFe/MgO through the Slonczewski-like torque due to the SHE[12–15]. However, the SHE-induced torque alone cannot directly drive the magnetostatically-preferred Bloch DWs[28] in these materials. We show experimentally and computationally that the DMI[1–7] provides the missing ingredient to explain current-induced DW motion in heavy-metal/ferromagnet/oxide systems[8–11] by stabilizing Néel DWs with a built-in chirality, such that the SHE alone drives them uniformly and with high efficiency. Engineering both the DW spin structure and the current-induced torque simply by selecting the materials adjacent to the ferromagnet presents unprecedented opportunities for designing current-controlled spintronic devices.

**Methods**

**Sample fabrication.** The stack structure of Pt/CoFe/MgO was Ta(3 nm)/Pt(3 nm)/Co$_{80}$Fe$_{20}$(0.6 nm)/MgO(1.8 nm)/Ta(2 nm), and that of Ta/CoFe/MgO was Ta(5 nm)/Co$_{80}$Fe$_{20}$(0.6 nm)/MgO(1.8 nm)/Ta(2 nm). Both were deposited on Si/SiO$_2$(50 µm) substrates. The metal layers were deposited by DC magnetron sputtering at 2 mTorr Ar (for Pt, 3 mTorr Ar), and MgO was RF sputtered at 3 mTorr Ar. The deposition rates were < 0.1 nm s$^-$ [1], calibrated with X-ray reflectivity. $Co_{80}Fe_{20}$ was chosen, instead of pure Co, to attain sufficient perpendicular magnetic anisotropy on both Ta and Pt underlayers. The bottom Ta(3 nm) layer in Pt/CoFe/MgO served as a seed layer to enhance perpendicular magnetic anisotropy and adhesion between Pt and the substrate. The Ta(2 nm) capping layer protected the MgO layer in each structure. Vibrating sample magnetometry on continuous films revealed full out-of-plane remanent magnetization and in-plane (hard-axis) saturation fields of ≈ 5 kOe for Pt/CoFe/MgO and ≈ 3 kOe for Ta/CoFe/MgO. The saturation magnetization was ≈700 $emu/cm^3$, approximately half of the bulk value, suggesting a magnetic dead layer due to roughness or oxidation. Both films exhibited weak DW pinning, with DW propagation at fields < 20 Oe.

The nanowires and Hall crosses were fabricated using electron beam lithography, magnetron sputtering, and liftoff. Electrical contacts consisting of Ta(2 nm)/Cu(100 nm) were placed with a second layer of electron beam lithography. To estimate the current density through these devices, current was assumed to flow only through the ultrathin CoFe layer and the adjacent heavy metal layer, so that the effective conductive thickness was 3.6 nm for Pt/CoFe/MgO and 5.6 nm for Ta/CoFe/MgO. We neglected current shunting in the bottom Ta seed layer in the Pt/CoFe/MgO, as sputtered Ta (beta phase) typically has a much higher resistivity than Pt. The resistance of the Ta/CoFe/MgO device was 3.5 times greater than the Pt/CoFe/MgO device, and the Ta layer was estimated to be 5 times more resistive than the Pt layer. Current shunting through the Ta capping layer, assumed to be oxidized, was also neglected.

**One dimensional model.** DW velocity was calculated using the standard one-dimensional (1D) model[29], which describes the DW in terms of two collective coordinates: position $X(t)$ and angle $\Phi(t)$, defined as the in-plane ($xy$) angle with respect to the positive $x$-axis. The current $\vec{j}_a = j_a \hat{x}$ is injected along the $x$-axis, and is positive along the positive $x$-axis.

(Note that the electron flow $j_e$ in the text is in the opposite direction with respect to $j_a$.) The 1D model including adiabatic and nonadiabatic STT, Slonczewski-like torque from the SHE, and the DMI[7] is given by

$$\begin{aligned}(1+\alpha^2)\frac{dX}{dt} &= \alpha\gamma_0\Delta H_z - \frac{\gamma_0\Delta H_K}{2}\sin(2\Phi) + (1+\alpha\beta)b_J \\ &+ \alpha\gamma_0\Delta\frac{\pi}{2}H_{SHE}\cos(\Phi) + \gamma_0\Delta\frac{\pi}{2}H_{DMI}\sin(\Phi) \\ &- \gamma_0\Delta\frac{\pi}{2}H_y\cos(\Phi) + \gamma_0\Delta\frac{\pi}{2}H_x\sin(\Phi)\end{aligned} \quad (1)$$

$$\begin{aligned}(1+\alpha^2)\frac{d\Phi}{dt} &= \gamma_0 H_z + \alpha\frac{\gamma_0 H_K}{2}\sin(2\Phi) + (\beta-\alpha)\frac{b_J}{\Delta} \\ &+ \gamma_0\frac{\pi}{2}H_{SHE}\cos(\Phi) - \alpha\gamma_0\frac{\pi}{2}H_{DMI}\sin(\Phi) \\ &+ \alpha\gamma_0\frac{\pi}{2}H_y\cos(\Phi) - \alpha\gamma_0\frac{\pi}{2}H_x\sin(\Phi)\end{aligned} \quad (2)$$

where $\Delta$ is the DW width, $H_K$ is the shape anisotropy field, $\alpha$ is the Gilbert damping, $\gamma_0$ is the gyromagnetic ratio, and $b_J$ is related to the adiabatic spin-transfer torque (STT), and is given by

$$b_J = \frac{\mu_B P}{eM_s}j_a, \quad (3)$$

and $\beta$ is the non-adiabatic parameter. In (3), $P$ is the spin polarization of the current, $\mu_B = 9.274\times 10^{-24}$ J/T is the Bohr magneton and $e = -1.6\times 10^{-19}$ C is the electron charge. The applied magnetic field has components ($H_x$, $H_y$, $H_z$). The Slonczewski-like torque from the SHE enters the 1D model equations via the effective field parameter

$$H_{SHE} = \frac{\hbar\theta_{SH}j_a}{2\mu_0 eM_s t_F} \quad (4)$$

where $\theta_{SH}$ is the spin Hall angle and $t_F$ is the thickness of the ferromagnetic layer. The effective field describing the Dzyaloshinskii-Moriya Interaction (DMI) is[7]

$$H_{DMI} = \frac{D}{\mu_0 M_s \Delta} \tag{5}$$

where $D$ is the DMI parameter. In the 1D model, the DMI enters as an effective field directed along the *x*-axis inside the DW[7], and promotes Néel DWs with internal magnetization oriented in either direction along the *x*-axis depending on the sign of $D$. The same chirality is therefore introduced for up-down and down-up DWs by using $D$ with opposite signs.

In order to qualitatively understand the experimental observations, archetypal parameters of a high perpendicular magnetic anisotropy (PMA) sample, with easy-axis along the *z*-axis, were considered: saturation magnetization $M_s = 3 \times 10^5$ A/m, exchange constant $A = 10^{-11}$ J/m, uniaxial anisotropy constant $K_u = 2 \times 10^5$ J/m$^3$, Gilbert damping $\alpha = 0.2$, polarization factor $P = 0.5$, non-adiabatic parameters $\beta = 0.4$, spin Hall angle $\theta_{SH} = 0.1$, and Dzyaloshinskii-Moriya constant $|D| = 0.5$ mJ/m$^2$. The 1DM inputs were chosen to be $\Delta = 8.32$ nm and $H_K = 12533$ A/m, which correspond to a ferromagnetic strip of rectangular cross-section $L_y \times L_z = 120$ nm$\times$3 nm explored in detail in Ref. 29. For these parameters and dimensions the static DW configuration in the absence of DMI was a Bloch DW magnetized along the *y*-axis, which was the initial DW configuration. Further details and model results are described in Supplementary Information.

**Acknowledgement**

This work was supported in part by the National Science Foundation under NSF-ECCS - 1128439. Technical support from David Bono is gratefully acknowledged. Devices were fabricated using instruments in the MIT Nanostructures Laboratory, the Scanning Electron-Beam Lithography facility at the Research Laboratory of Electronics, and the Center for Materials Science and Engineering at MIT. S.E. acknowledges financial support by the NSF Graduate Research Fellowship Program. The work by E. M. was supported by projects MAT2011-28532-C03-01 from the Spanish government and SA163A12 from Junta de Castilla y Leon.

**Author Contributions**

G.S.D.B. proposed and supervised the study. S.E. and G.S.D.B. designed the experiments. S.E. and U.B. built the measurement setups with assistance from G.S.D.B. S.-M.A. developed and deposited the Ta/CoFe/MgO and Pt/CoFe/MgO films. S.E. carried out the lithographic steps and performed all measurements. E.M. performed the modeling and wrote the corresponding text. S.E. analyzed the data. S.E. and G.S.D.B. wrote the manuscript with assistance from U.B. All authors discussed the results and commented on the manuscript.

**Additional Information**

Correspondence and requests for materials should be addressed to G.S.D.B.

**Competing Financial Interests**

The authors declare no competing financial interests.

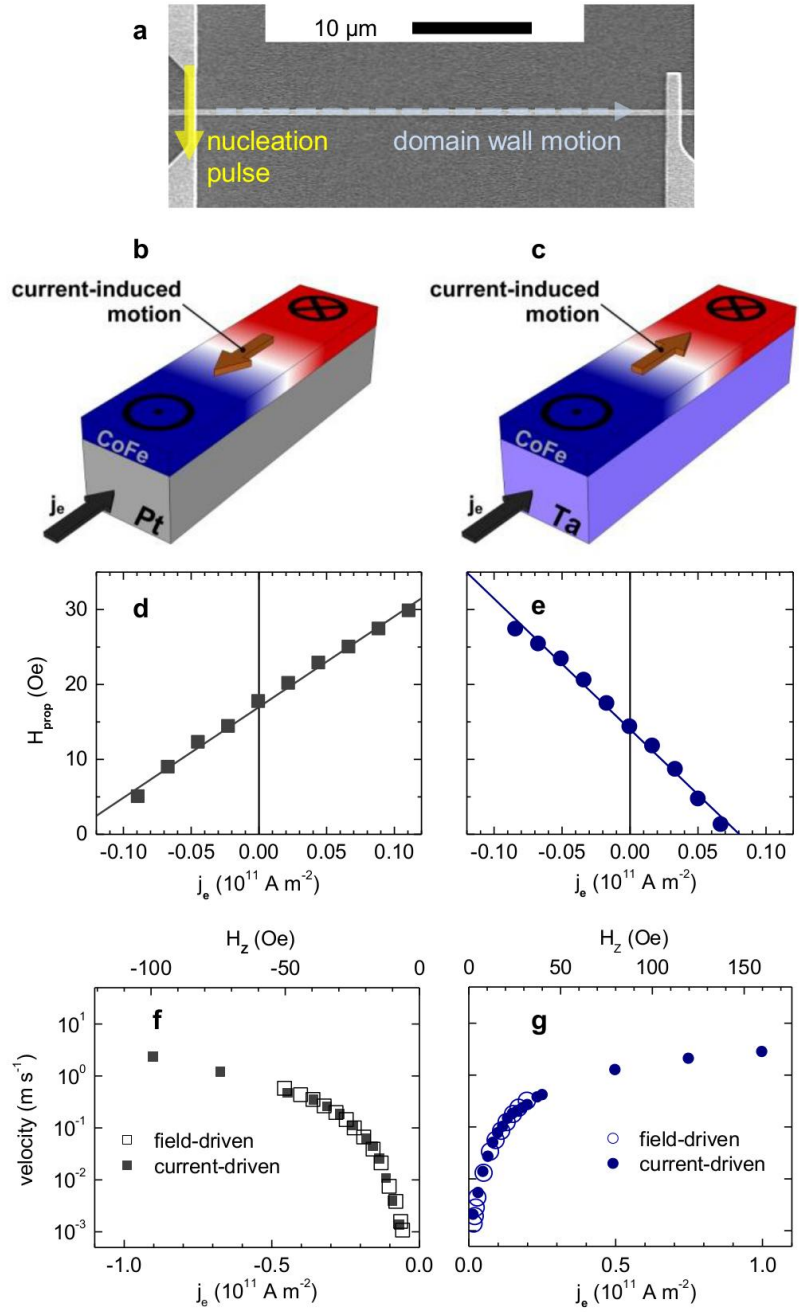

Fig. 1

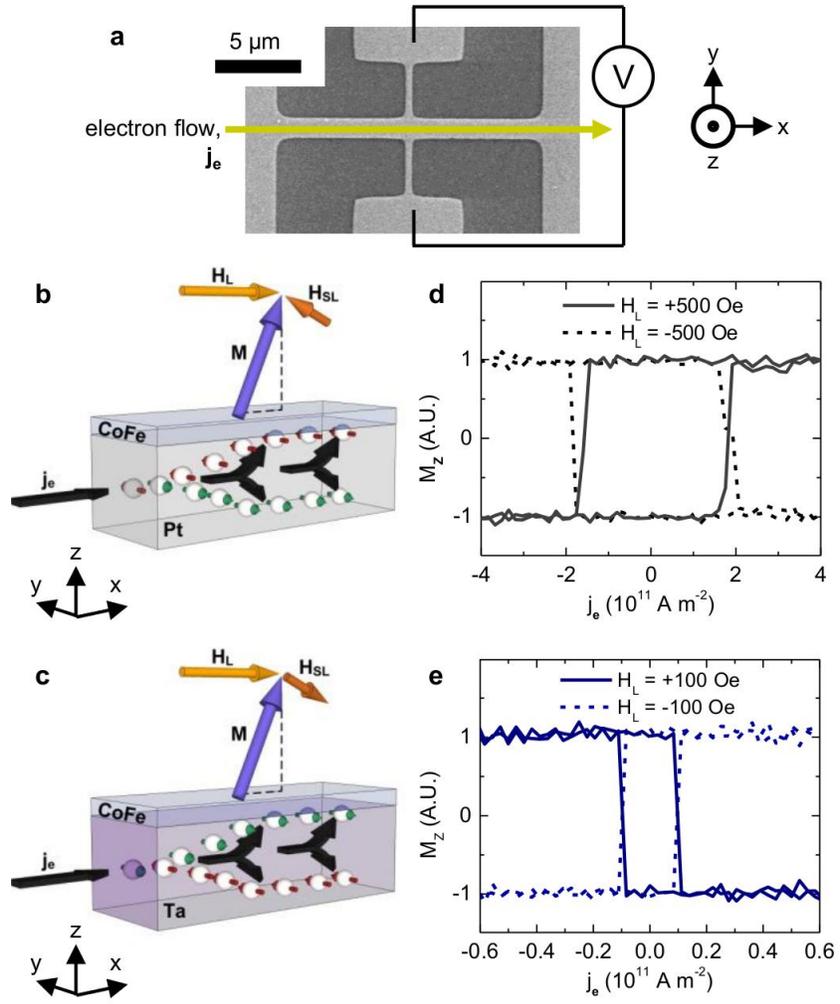

Fig. 2

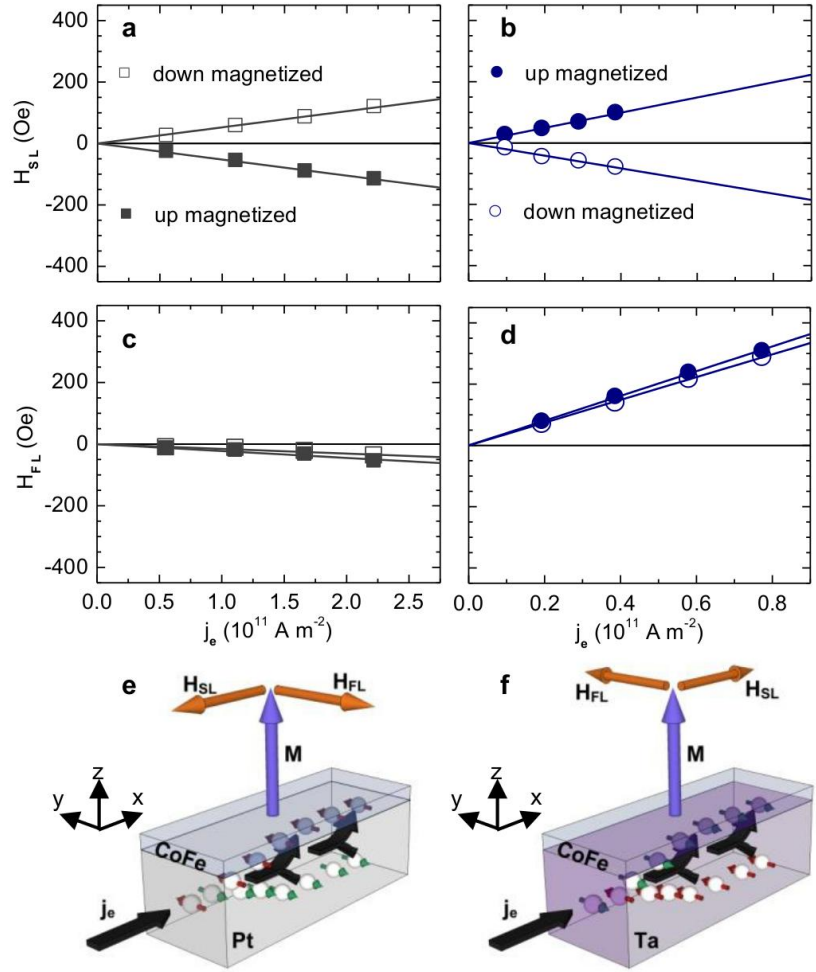

Fig. 3

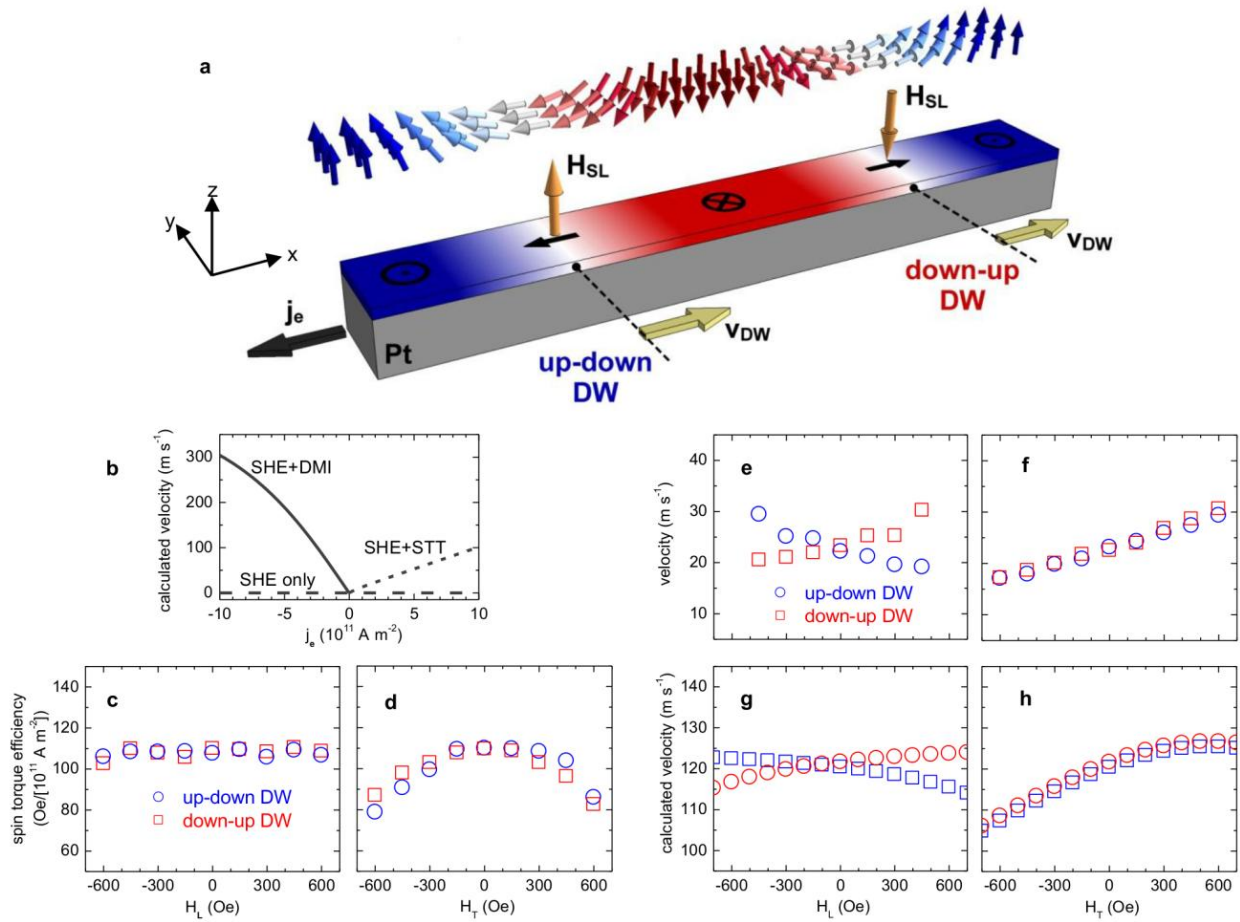

Fig. 4

**Figure 1 | Effect of current on domain wall motion. a,** Scanning electron micrograph of the nanowire. The current pulse on the left nucleates a domain wall, which is then propagated to the right by current or applied out-of-plane field. **b, c,** Illustrations of the direction of current-driven domain wall motion in the Pt/CoFe/MgO (**b**) and Ta/CoFe/MgO (**c**) nanowires. Electron current $j_e$ is defined positive when conduction electrons flow away from the nucleation line, from left to right in the micrograph (**a**). **d, e,** Domain wall propagation field $H_{\text{prop}}$ as a function of driving electron current density $j_e$ for Pt/CoFe/MgO (**d**) and Ta/CoFe/MgO (**e**). The slope of the linear fit extracts the spin-torque efficiency for each structure. **f, g,** Domain wall velocity as a function of $j_e$ and applied out-of-plane field $H_z$ for Ta/CoFe/MgO (**f**) and Pt/CoFe/MgO (**g**). The field-driven data are scaled by a field-to-current ratio (see text) so that they are directly on top of the current-driven data.

**Figure 2 | Current-induced switching under a constant in-plane longitudinal field. a,** Scanning electron micrograph of a Hall cross. **b, c,** Illustrations of Pt/CoFe/MgO (**b**) and Ta/CoFe/MgO (**c**) in the up magnetization state with the injected electron current and applied longitudinal field $H_L$ in the +$x$ direction. Because of the combination of the current-induced Slonczewski-like torque (producing an effective field $H_{\text{SL}}$) and the applied longitudinal field, up magnetization is stable in Pt/CoFe/MgO whereas it is unstable in Ta/CoFe/MgO. **d, e,** Out-of-plane magnetization $M_z$ (normalized anomalous Hall signal) as a function of electron current density $j_e$ under a constant $H_L$ in Pt/CoFe/MgO (**d**) and Ta/CoFe/MgO (**e**). The magnitude of $H_L$ is 500 Oe for Pt/CoFe/MgO (**d**) and 100 Oe for Ta/CoFe/MgO (**e**). When $H_L$ is reversed from +$x$ (solid line) to –$x$ (dotted line), the stable magnetization direction under a given current polarity reverses.

**Figure 3 | Current-induced effective fields. a, b,** Current-induced effective longitudinal field $H_{SL}$, arising directly from the Slonczewski-like torque, as a function of electron current density $j_e$ (from AC excitation current amplitude) in Pt/CoFe/MgO (**a**) and Ta/CoFe/MgO (**b**). **c, d,** Current-induced effective transverse field $H_{FL}$ as a function of $j_e$ in Pt/CoFe/MgO (**c**) and Ta/CoFe/MgO (**d**). **e, f,** Illustration of the directions of the current-induced effective fields $H_{SL}$ and $H_{FL}$ in Pt/CoFe/MgO (**e**) and Ta/CoFe/MgO (**f**), when the magnetization is up and the electron flow is in the $+x$ direction.

**Figure 4 | Current-driven dynamics of chiral Néel domain walls. a,** Illustration of left-handed chiral Néel domain walls in Pt/CoFe/MgO. The effective field $H_{SL}$ from the Slonczewski-like torque moves adjacent up-down and down-up domains with velocity $v_{DW}$ in the same direction against electron flow $j_e$. **b,** Domain wall velocity as a function of electron current density $j_e$, calculated using the one-dimensional model, with the spin Hall effect only (SHE only), the spin Hall effect and spin-transfer torque (SHE+STT), and the spin Hall effect and the Dzyaloshinskii-Moriya interaction (SHE+DMI). The parameters used in this calculation are in the Methods section. **c, d,** Spin-torque efficiency for domain wall motion in Pt/CoFe/MgO under applied longitudinal field $H_L$ (**c**) and transverse field $H_T$ (**d**). **e, f,** Domain wall velocity at a constant current $j_e = -3.0 \times 10^{11}$ A m$^{-2}$ as a function of $H_L$ (**e**) and $H_T$ (**f**). **g, h,** Calculated domain wall velocity at $j_e = -3.0 \times 10^{11}$ A m$^{-2}$ as a function of $H_L$ (**g**) and $H_T$ (**h**) using the one-dimensional model.

# Current-driven dynamics of chiral ferromagnetic domain walls

- Supplementary Information -

Satoru Emori[1], Uwe Bauer[1], Sung-Min Ahn[1], Eduardo Martinez[2], Geoffrey S. D. Beach[1*]

[1] Department of Materials Science and Engineering, Massachusetts Institute of Technology,

Cambridge, Massachusetts 02139, USA

[2] Dpto. Física Aplicada. Universidad de Salamanca,

Plaza de los Caidos s/n E-38008, Salamanca, Spain

## I. Measurement of the domain wall propagation field $H_{prop}$

A lithographically patterned 40-μm long, 500-nm wide magnetic nanowire is shown in Fig. S1a. Magnetization switching was measured using our custom scanning magneto-optical Kerr effect (MOKE) system[1] with a ~3 μm focused laser beam placed at a fixed position midway along the nanowire. For each value of electron current density $j_e$ injected through the nanowire, MOKE hysteresis loops were obtained under an applied out-of-plane field $H_z$ with a triangular waveform of amplitude 150 Oe and frequency 12.5 Hz. After saturating the magnetic nanowire (e.g. uniformly magnetized down), a reverse domain (e.g. up) was nucleated using a 25-ns, 50-mA nucleation pulse just after the field zero-crossing. With the increasing $H_z$ expanding the reverse domain, a domain wall (DW) propagated away from the nucleation line, and magnetization switching was detected by MOKE as the DW passed through the laser spot. After saturating the magnetic nanowire by domain expansion, $H_z$ was swept in the other direction. For this side of field sweep, no DW was initialized so that magnetization switching occurred through domain nucleation at random locations in the nanowire. At a sufficiently large $H_z$, the magnetization was saturated again in the initial direction. This measurement cycle was repeated at least 50 times and an averaged hysteresis loop was obtained to attain a sufficient signal-to-noise ratio and take into account the stochasticity of magnetization switching. The DW propagation field $H_{prop}$ was taken as the field at which the normalized MOKE signal (Fig. S1b) crossed zero. As seen on the positive side of $H_z$ in Fig. S1b, we observed a clear shift of $H_{prop}$ even at small currents < 10 µA ($|j_e| < 10^{10}$ A m$^{-2}$), resulting in a linear correlation between $H_{prop}$ and driving current as shown in Fig. 1 of the Letter. By contrast, no systematic variation in the nucleation field (switching field in the absence of an initialized DW) was observed with respect to current, as shown on the negative field side of $H_z$ in Fig. S1b.

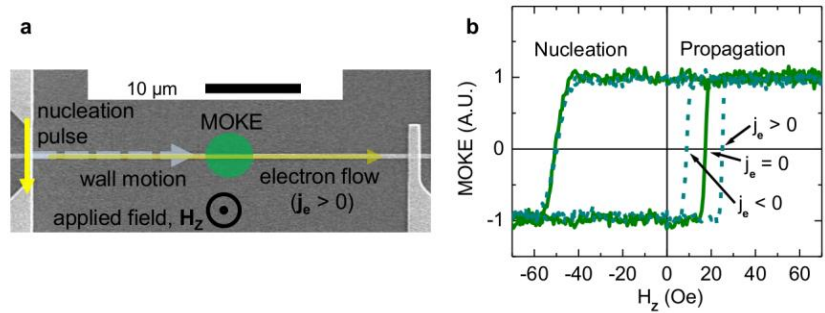

**Figure S1** | **a,** Schematic of the domain wall propagation field measurement superposed on a scanning electron micrograph of a nanowire. The focused laser spot is placed halfway along the nanowire for magneto-opical Kerr effect (MOKE) measurements. **b,** Exemplary MOKE hysteresis loops on a Pt/CoFe/MgO device at under polarities of current densities (here, $|j_e|$ = 0.07×10$^{11}$ A m$^{-2}$). A domain wall is initialized only on the rising side of the positive applied field, so that here the positive coercive field is the DW propagation field while the negative coercive field is the reverse domain nucleation field.

Some data of DW propagation field were obtained with a high-impedance DC source outputting the current to drive DWs. However, initialized DWs were occasionally pinned around the initialization line at larger magnitudes of driving current. We speculate that the voltage drop across the thin MgO layer might have locally modified magnetic anisotropy in the vicinity of the initialization line, through an effect similar to what was observed in Ref 2. To circumvent any issues associated with this pinning effect, we conducted most propagation field measurements in the following way: With the nanowire saturated (e.g. fully magnetized in the down direction), an out-of-plane field was ramped to ~20-30 Oe (less than the nucleation field) in 100 μs. While this field was maintained, a reverse domain (e.g. up) was nucleated with a current pulse in the initialization line. The field was maintained for another 50 μs to de-pin and drive the DW away from the initialization line by ~5 μm. After 1 ms, we then injected a current through the magnetic nanowire (output by a function generator with a rise-time of ~10 ns) and began sweeping the out-of-plane field at the same time. The difference in the extracted spin-torque efficiency depending on the measurement method was as much as ~20 %, but the polarity of the spin-torque efficiency did not change with the measurement method.

The DW motion data presented in Fig. 1 of the Letter were conducted at a constant substrate temperature of 308 K, maintained to within ±0.1 K with a thermoelectric module. The DW motion data under applied in-plane fields shown in Fig. 4 were not conducted on the thermoelectric module; the substrate temperature was within 295-300 K.

## II. Current-driven domain wall motion for opposite senses of magnetization

A recent report by Haazen *et al.*[3] on all-metal asymmetric Pt/Co/Pt stacks shows that current effectively promoted domain expansion or contraction, i.e. up-down and down-up DWs moved in opposite directions with respect to current under an in-plane longitudinal field. Current could displace DWs only under a sufficient in-plane longitudinal field that locked the Néel wall configuration. Adjacent up-down and down-up Néel walls were expected to have opposite chiralities, with their internal magnetic moments oriented parallel to the longitudinal field, which would result in opposite directions of motion with respect to current. When the direction of the longitudinal field was reversed, the direction of motion for each DW was reversed. These results of current-driven DW motion in Pt/Co/Pt were attributed to the Slonczewski-like torque from the spin Hall effect.

By contrast, recent studies on Pt/Co/AlOx[4,5] and Pt/Co/GdOx[6] have reported up-down and down-up DWs moving in the same direction under current. We measured several different nanowires of Pt/CoFe/MgO and Ta/CoFe/MgO, and current-driven DW motion was also unidirectional in those devices irrespective of the magnetization sense across the DW. In particular, for both senses of magnetization across the DW ("Up-Down" and "Down-Up" DWs), the same trend was observed (Fig. S2): With electron flow in the same direction as field-induced DW motion, the propagation field increased in Pt/CoFe/MgO and decreased in Ta/CoFe/MgO. In other words, the motion of both Up-Down and Down-Up DWs was hindered in the direction of electron flow in Pt/CoFe/MgO and assisted in Ta/CoFe/MgO. The unidirectional motion of Up-Down and Down-Up DWs requires a Néel configuration with fixed chirality as discussed in the Letter and in Sections IX and X of this Supplementary document.

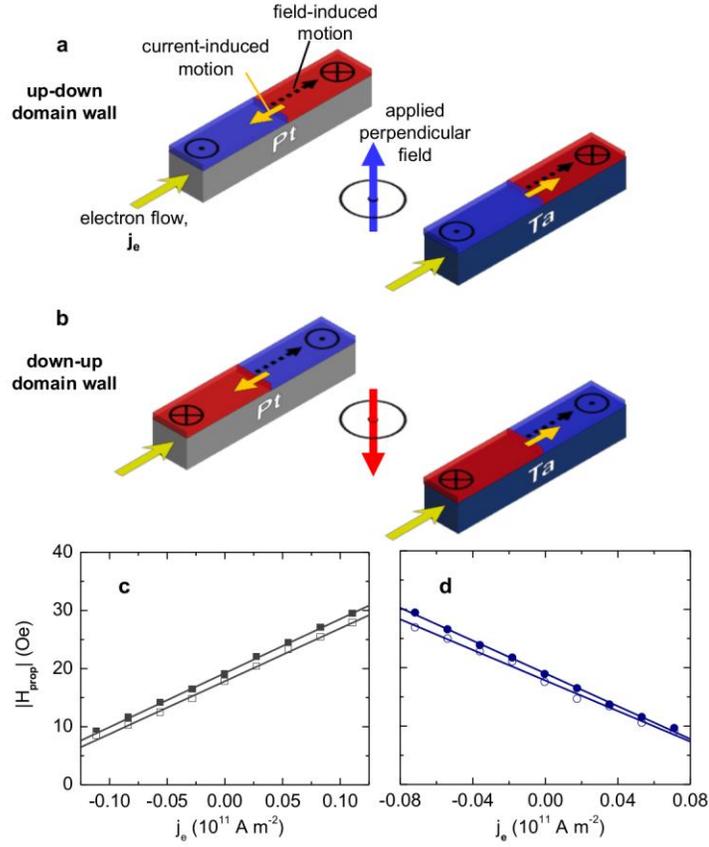

**Figure S2 | a,b** Direction of current-induced motion in Pt/CoFe/MgO and Ta/CoFe/MgO for an up-down domain wall (up domain expanding under an upward applied field) (**a**) and a down-up domain wall (down domain expanding under a downward applied field) (**b**). **c,d** Domain wall propagation field $H_{prop}$ as function of driving electron current density $j_e$ exhibiting the same trend for Up-Down (filled symbols) and Down-Up domain walls (empty symbols) in Pt/CoFe/MgO (**c**) and Ta/CoFe/MgO (**d**).

### III. Velocity measurements of current-driven DWs

Current-driven DW velocity measurements were carried out by first driving a DW away from the nucleation line with the out-of-plane field pulse (as described in Section I) and then injecting the current to drive the DW. The field-driven DW velocity was measured by first ramping the field to the setpoint driving level (because of the slow rise-time of the magnet ~100 µs) and then initializing a DW, which was subsequently driven to the other end of the nanowire by the constant driving field. The DW arrival time was extracted from MOKE signal versus time averaged over at least 50 measurement cycles, and the velocity was obtained by linearly fitting the plot of the arrival time at several positions along the nanowire, as shown in Fig. S3.

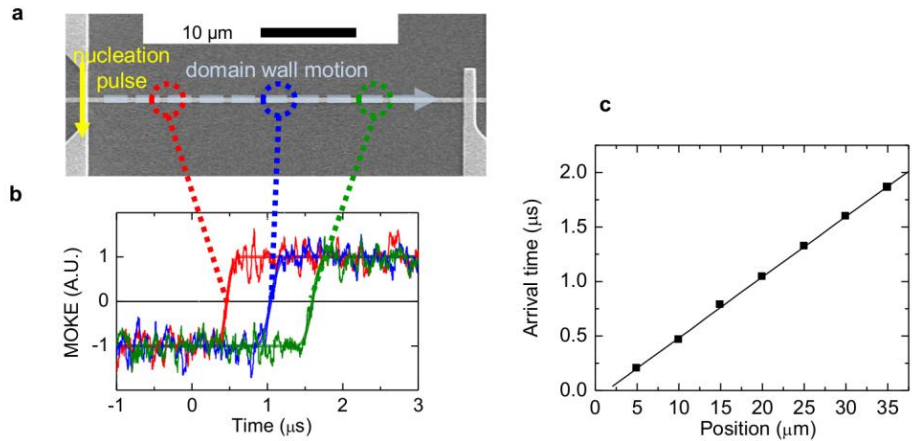

**Figure S3 | a,** Scanning electron micrograph of a 500-nm wide nanowire. The current pulse on the left initializes a domain wall, which is then driven to the right by an applied field or injected current. **b,** Magneto-optical Kerr effect signal showing magnetization switching due to domain wall propagation at different positions along the nanowire. **c,** Domain wall arrival time plotted against position along the nanowire. The domain wall velocity is extracted by linear fit.

## IV. Domain wall velocity at higher current densities

In Fig. S4, we present the velocity of DWs up to electron current densities $j_e$ larger than the range shown in Fig. 1 of the Letter. Interestingly, with $|j_e| > 1\times10^{11}$ A/m$^2$, the DW velocity is higher in Pt/CoFe/MgO than in Ta/CoFe/MgO, despite the smaller spin-torque efficiency and spin Hall angle exhibited by Pt/CoFe/MgO. This paradox suggests that the DW velocity is not necessarily an accurate metric of the current-induced torque, and that caution must be used when relying on velocity data alone to compare the current-induced torques in different materials. The origin of the paradox may be the difference in magnetization damping for these two structures: Pt-based magnetic thin films typically exhibit much stronger damping (with damping parameters ~0.1 or greater)[7–9] than Ta-based ones (~0.01)[10,11]. The higher velocity in Pt/CoFe/MgO is then explained if the current-driven DW mobility scales with the damping parameter. Further computational and experimental studies will elucidate the effect of damping on chiral Néel DWs driven by the Slonczewski-like torque.

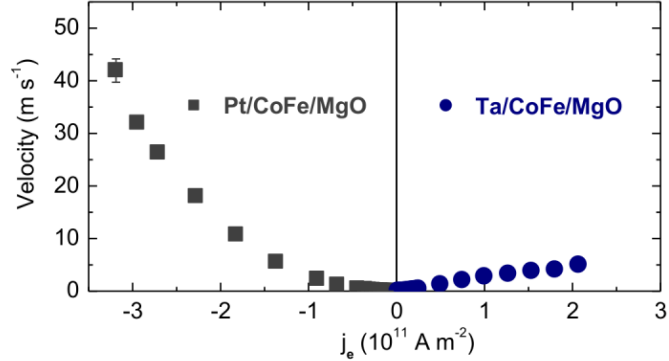

**Figure S4 |** Domain wall velocity as a function of electron current density $j_e$ in Pt/CoFe/MgO and Ta/CoFe/MgO.

## V. Measurement of current-induced switching

Magnetization switching was induced by 250-ms long current pulses injected in the longitudinal (*x*) direction of the cross (Fig. 2a). 250 ms after the current pulse was turned off, the out-of-plane component of the magnetization was measured from the anomalous Hall voltage using a 400-Hz low-amplitude (~$10^9$ A m$^{-2}$) AC sense current and a lock-in amplifier. The out-of-plane magnetization was measured successively in this fashion by stepping through a range of current pulse amplitudes, producing hysteresis loops as shown in Fig. 2 of the Letter.

## VI. Measurement of current-induced effective longitudinal field

Figs. S5a,b illustrate the measurement setup for determining the magnitude and direction of the effective in-plane longitudinal field $H_{SL}$ in a uniformly magnetized Hall cross. Two lock-in amplifiers were used to measure the first and second harmonics in the anomalous Hall voltage signal simultaneously. The frequency of the AC (sinusoidal) excitation current $\omega/2\pi$ was 80 Hz. The measured second harmonic was set 90 degrees out of phase with respect to the first harmonic (see equations below). The applied in-plane longitudinal field $H_L$ from an electromagnet was swept from +600 Oe to -600 Oe and then back to +600 Oe quasistatically. An air-coil was placed directly beneath the sample substrate to apply a constant out-of-plane field of 32 Oe to keep the magnetization from switching.

The work by Kim et al.[12] describes the derivation of the model used in our study to extract $H_{SL}$ produced by the Slonczewski-like torque. When the Hall cross is uniformly magnetized along the out-of-plane (*z*) direction, the AC excitation current generates an effective field along the longitudinal (*x*) axis, which modulates $H_L$, i.e.

$$H_L^{total} = H_L + H_{SL} \sin \omega t.$$

This sinusoidal modulation of the longitudinal field results in an anomalous Hall voltage $V_{AH}$ of the form

$$V_{AH} = V_{DC} + V_\omega \sin \omega t + V_{2\omega} \cos 2\omega t,$$

where $V_{DC}$ is the component that does not depend on the frequency of the AC excitation, $V_\omega$ is the in-phase first harmonic, and $V_{2\omega}$ is the out-of-phase second harmonic. The expressions for $V_\omega$ and $V_{2\omega}$ are

$$V_\omega = \pm \Delta R_{AH} I_O \left(1 - \frac{H_L^2}{2D^2}\right) \text{ and } V_{2\omega} = \pm \frac{\Delta R_{AH} I_O}{2D^2} H_L H_{SL},$$

where $\Delta R_{AH}$ is the difference in the anomalous Hall resistance between the up and down magnetized states and $I_O$ is the amplitude of the AC excitation current. $D$ is a constant defined by the out-of-plane uniaxial anisotropy constant $K_U$, saturation magnetization $M_S$, and applied out-of-plane field $H_z$: $D \equiv (2K_U/M_S) - 4\pi M_S + H_z$. The $\pm$ in front of the equations correspond to the up (+ z) magnetized and down (- z) magnetized states, respectively. The ratio of the first derivative of $V_{2\omega}$ to the second derivative of $V_\omega$ is taken to solve for $H_{SL}$:

$$H_{SL} = -2\left(\frac{dV_{2\omega}}{dH_L}\right) \bigg/ \left(\frac{d^2 V_\omega}{dH_L^2}\right).$$

In the above expression, the sign of $H_{SL}$ indicates the direction of $H_{SL}$ (+x or –x) when conventional current is positive, i.e. in the + x-direction. In the Letter, we discuss the directions of $H_{SL}$ when *electron flow* is in the + x-direction (conventional current in the –x direction). Therefore, the appropriate expression for $H_{SL}$ in the Letter has the opposite sign.

$$H_{SL} = 2\left(\frac{dV_{2\omega}}{dH_L}\right) \bigg/ \left(\frac{d^2 V_\omega}{dH_L^2}\right).$$

There are a few key assumptions that go into this model[12]. (1) The Hall cross is uniformly magnetized. (2) The tilting $\theta$ of the out-of-plane magnetization is small (small angle approximation $\sin\theta \approx \theta$). (3) The magnetization is not tilted in the transverse direction, although in general current can generate a torque that pulls the magnetization in the transverse direction as well.

Exemplary data of $V_\omega$ and $V_{2\omega}$ are shown in Figs. S5c-j. As expected, $V_\omega$ data for both Pt/CoFe/MgO and Ta/CoFe/MgO vary quadratically with $H_L$ as the magnetization tilts away from the perpendicular axis. We observed different offsets in $V_\omega$ in various devices due to a voltage drop from < 0.1% of excitation current flowing in the transverse direction of the Hall cross (≈ 17 kΩ for Ta/CoFe/MgO, ≈ 5 kΩ for Pt/CoFe/MgO). The maximum magnetization tilting at 600 Oe of applied longitudinal field is ≈ 20 ° (0.35 rad) in Ta/CoFe/MgO and ≈ 10 ° (0.17 rad) in Pt/CoFe/MgO. The variation of $V_{2\omega}$ is linear with respect to longitudinal field in Pt/CoFe/MgO, consistent with the model described above.

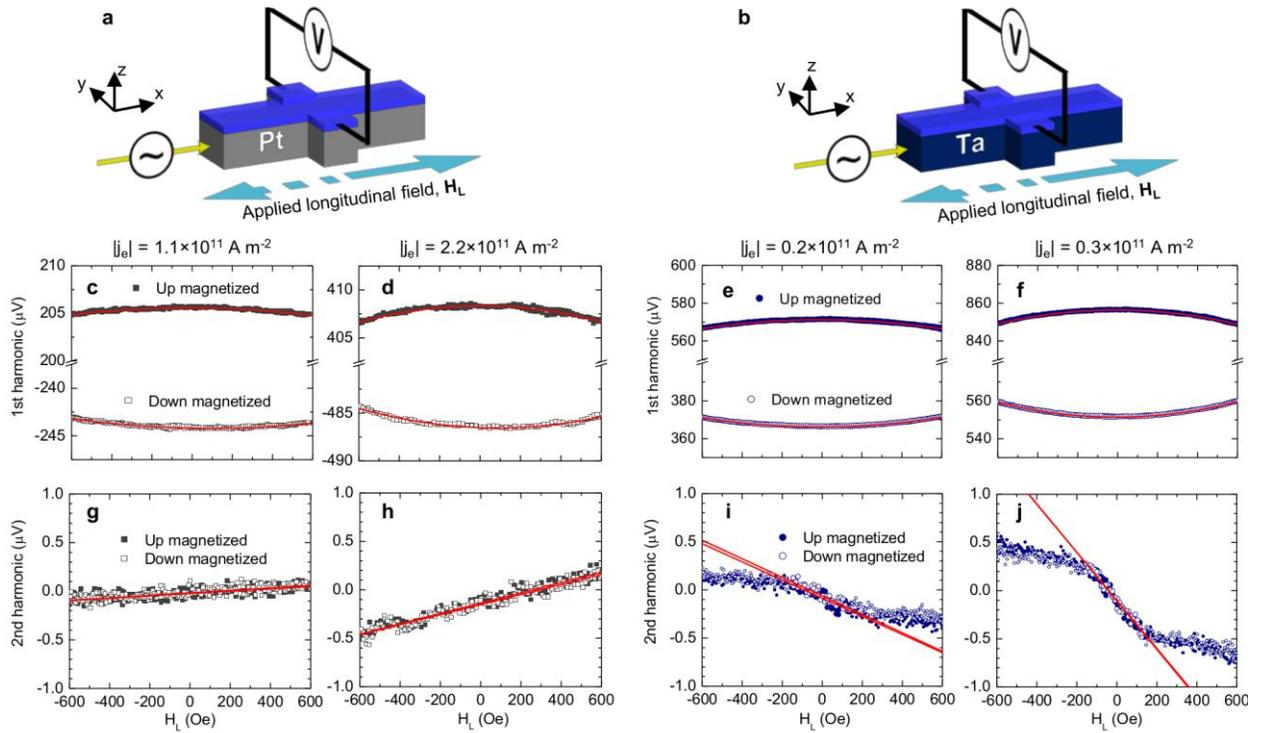

**Figure S5 | a,b** Schematics of the anomalous Hall voltage measurement to determine the effective longitudinal field $H_{SL}$ arising from the Slonczewski-like torque in Pt/CoFe/MgO (**a**) and Ta/CoFe/MgO (**b**). **c-f,** First harmonic of the anomalous Hall voltage as a function of applied in-

plane longitudinal field $H_L$ in Pt/CoFe/MgO (**c,d**) and Ta/CoFe/MgO (**e,f**). **g-j,** Second harmonic of the anomalous Hall voltage as a function of $H_L$ in Pt/CoFe/MgO (**g,h**) and Ta/CoFe/MgO (**i,j**).

However, for Ta/CoFe/MgO, different linear slopes of $V_{2\omega}$ were observed, with a larger slope for $|H_L| <$ 100-200 Oe and a reduced slope for $|H_L| >$ 100-200 Oe. We observed this behavior in all measured Ta/CoFe/MgO Hall crosses. To explore the origin of this behavior, $V_\omega$ and $V_{2\omega}$ were measured under different constant bias out-of-plane fields as shown in Fig. S6. For a small excitation current (Figs. S6a,c,e), magnetization switching was observed and $V_{2\omega}$ diverged around $H_L$ = 100 Oe under zero out-of-plane field. The switching and divergence were suppressed under nonzero out-of-plane fields, although the transition in the $V_{2\omega}$ slope remained. At an even higher excitation current amplitude (Fig. S4b,d,f), diverging features at $H_L \approx$ 100 Oe could no longer be suppressed by out-of-plane field. Although we do not have a complete explanation for the origin of these features, we can attribute the transition in the $V_{2\omega}$ slope appearing at $H_L \approx$100-200 Oe to the formation of a multi-domain state (i.e. deviation from uniform magnetization) or the onset of instability in the magnetization direction. For example, the combination of a sufficient longitudinal field and excitation current could possibly destabilize the magnetization around the edges of the Hall cross, where perpendicular magnetic anisotropy might be weaker. In extracting $H_{SL}$, we took the steeper slope of $V_{2\omega}$ around zero longitudinal field, where the out-of-plane magnetization was expected to be uniform and $H_L$ is not large enough to drive the instability. We also limited the range of excitation current amplitude (Fig. 4 main text) so that both $V_\omega$ and $V_{2\omega}$ data could be fit quadratically and linearly, respectively, over a sufficient range of $H_L$.

The estimated Hall angle in Ta based on $H_{SL} \approx -200$ Oe/$10^{11}$ A m$^{-2}$ presented in the main text (Fig. 4) is extraordinarily large at -0.25. We further note that an even larger $H_{SL}$ of around -400 Oe/$10^{11}$ A m$^{-2}$ was observed for another Ta/CoFe/MgO Hall cross. This suggests that our results deviate from the ideal extraction of $H_{SL}$ described in Ref. 12, likely due to the instability in the magnetically soft Ta/CoFe/MgO (with relatively weak perpendicular magnetic anisotropy). Regardless of the deviation in the magnitude, the slopes of $V_{2\omega}$ are clearly opposite in Pt/CoFe/MgO and Ta/CoFe/MgO. We emphasize that the directions of the Slonczewski-like torque, and hence the resulting effective fields, in Pt/CoFe/MgO and Ta/CoFe/MgO are opposite.

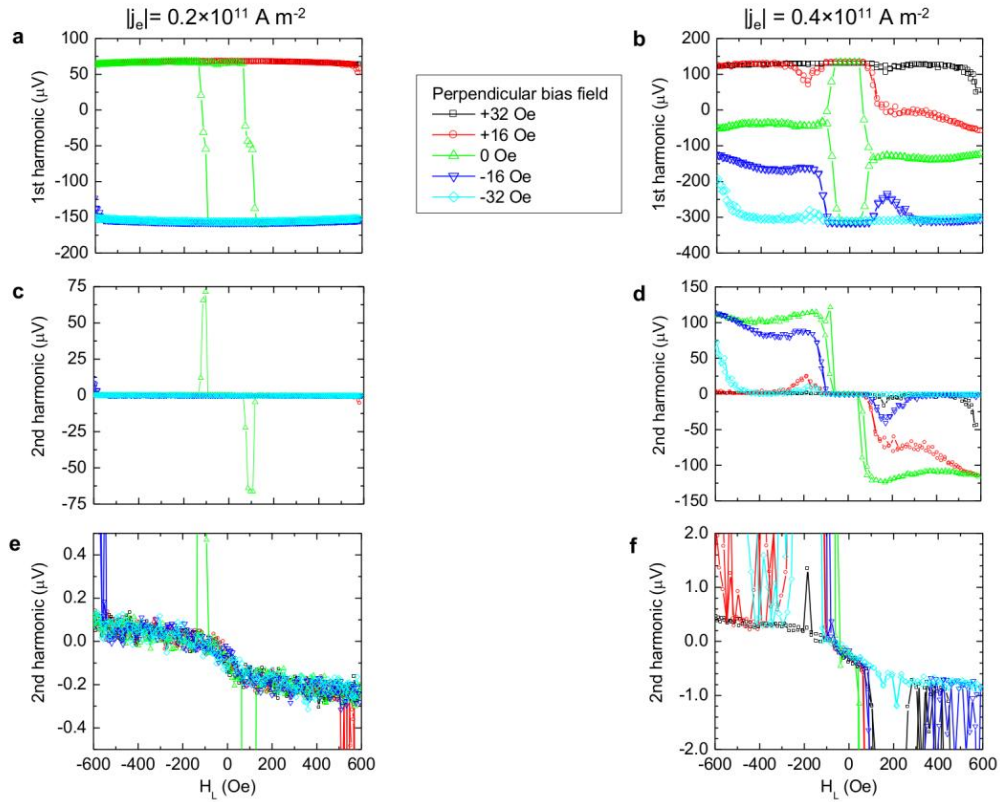

**Figure S6 | a,b,** First harmonic of the anomalous Hall voltage in Ta/CoFe/MgO at several different out-of-plane bias magnetic fields. **c-f,** Second harmonic of the anomalous Hall voltage

in Ta/CoFe/MgO at several different out-of-plane bias magnetic fields. Close-ups of the slope transition are also shown (**e,f**).

## VII. Measurement of current-induced effective transverse field

The measurement of the current-induced effective transverse field $H_{FL}$ was carried out in the same fashion as the measurement of $H_{SL}$, except with the applied in-plane field $H_T$ in the transverse (*y*) direction. Based on Ref. 12, the expressions for the first and second harmonics of the anomalous Hall voltage are

$$V_\omega = \pm \Delta R_{AH} I_O \left(1 - \frac{H_T^2}{2D^2}\right) \text{ and } V_{2\omega} = \pm \frac{\Delta R_{AH} I_O}{2D^2} H_T H_{FL}.$$

The resulting expression for $H_{FL}$ is then

$$H_{FL} = -2 \left(\frac{dV_{2\omega}}{dH_T}\right) \bigg/ \left(\frac{d^2 V_\omega}{dH_T^2}\right).$$

In the Letter, we again discuss the directions of $H_{FL}$ when *electron flow* is along +*x*, so the appropriate expression for $H_{FL}$ is

$$H_{FL} = 2 \left(\frac{dV_{2\omega}}{dH_T}\right) \bigg/ \left(\frac{d^2 V_\omega}{dH_T^2}\right).$$

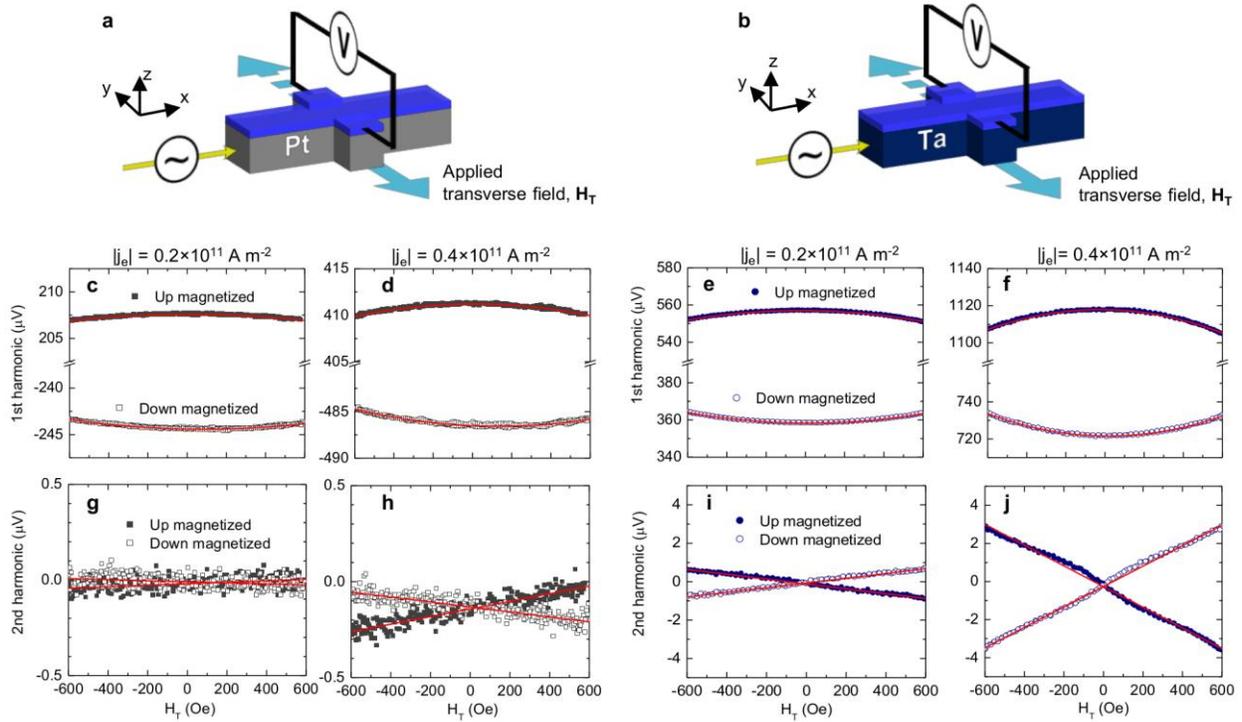

**Figure S7 | a,b,** Schematics of the anomalous Hall voltage measurement to determine the effective transverse field $H_{FL}$ in Pt/CoFe/MgO (**a**) and Ta/CoFe/MgO (**b**). **c-f,** First harmonic of the anomalous Hall voltage as a function of applied in-plane transverse field $H_T$ in Pt/CoFe/MgO (**c,d**) and Ta/CoFe/MgO (**e,f**). **g-j,** Second harmonic of the anomalous Hall voltage as a function of $H_T$ in Pt/CoFe/MgO (**g,h**) and Ta/CoFe/MgO (**i,j**).

The opposite directions of $H_{FL}$ observed in Pt/CoFe/MgO and Ta/CoFe/MgO may arise from the Rashba effect, although we cannot conclude whether there is any difference in Pt/CoFe/MgO and Ta/CoFe/MgO that may induce opposite polarities of the Rashba field. A more straightforward explanation, as pointed out in the Letter and in Ref. 13, is that the observed $H_{FL}$ may be an artifact of the Slonczewski-like torque in the presence of an applied transverse field. Further work is required to verify the origin of the large $H_{FL}$ observed for Ta/CoFe/MgO.

## VIII. Current-driven DW motion in various out-of-plane magnetized ultrathin layered structures

In this present work, we discuss the anomalously efficient current-driven DW dynamics in heavy-metal/ferromagnet/oxide structures. Similar DW dynamics have been observed in Pt(15 Å)/Co(3 Å)/Ni(7 Å)/Co(1.5 Å) capped by a TaN overlayer[14]. This suggests that insulating materials other than oxides can be used for efficient motion of chiral DWs propelled by the Slonczewski-like torque from the spin Hall effect. Furthermore, Ref. 15 reports multiple DWs moving uniformly against electron flow in an ultrathin structure of Pt/Co/Pt with different thicknesses for Pt underlayer and overlayer. Many studies have also shown current-induced DW displacement in Co/Pt multilayers, whereas others have reported no such current-induced effect other than Joule heating. The recent study by Haazen *et al.*[3] reports DW displacement in asymmetric Pt/Co/Pt due to the spin-Hall effect only when a finite longitudinal magnetic field is applied, in contrast to the report[16] of systematic current-induced DW displacement in similar structures without any applied longitudinal field.

This wide discrepancy may arise from the requirement for current-induced DW motion governed by the spin Hall effect: asymmetry in the interfaces is necessary to stabilize a Néel wall through the Dzyaloshinskii-Moriya interaction. Recent studies[17,18] have been shown that, in Pt/Co/Pt, the magnetic anisotropy arising from the bottom Pt/Co interface is significantly stronger than that from the top Co/Pt interface. On the other hand, the relative strengths of interfacial anisotropy may vary in structures with different underlayers, deposition conditions, and post-deposition treatment processes (e.g. ion irradiation). Such subtle interfacial differences could have resulted in the disparate efficiencies of current-induced DW displacement in Co/Pt-based structures.

More generally, the interface between the ferromagnet and the overlayer (in addition to the interface between the ferromagnet and the underlayer) may also influence the efficiency of current-driven DW motion. For example, according to our recent work on Pt/Co/GdOx[6], disrupting the Co/GdOx interface with a 4-Å thick Pt layer nearly halved the spin-torque efficiency, compared to the equivalent structure with an uninterrupted Co/GdOx interface. This Pt dusting layer was likely not continuous, and its thickness was significantly smaller than the spin diffusion length in Pt ($\approx$ 1.4 nm)[19]. Therefore, the Pt dusting layer did not generate a Slonczewski-like torque that counteracted the torque from the Pt underlayer; the reduction in the spin-torque efficiency instead was caused directly by the altered Co/GdOx interface. Although the Dzyaloshinskii-Moriya interaction (DMI) in an ultrathin ferromagnet has been reported to arise from its interface with a nonmagnetic heavy metal[20,21], it is possible that a ferromagnet/oxide (or ferromagnet/insulator) interface contributes to the DMI as well.

## IX. Micromagnetic description of torques

The general form of the Landau-Lifshitz-Gilbert equation including the spin-transfer torque (STT) and the Slonczewski-like torque (SLT) from the spin Hall effect (SHE) is

$$\frac{\partial \vec{m}}{\partial t} = -\gamma_0 \vec{m} \times \vec{H}_{eff} + \alpha\, \vec{m} \times \frac{\partial \vec{m}}{\partial t} + \vec{\tau}_{STT} + \vec{\tau}_{SLT}^{SHE}, \qquad (1)$$

where $\vec{m} = \vec{M}/M_s$ is the normalized local magnetization ($|\vec{m}|=1$), and $\gamma_0 = 2.21 \times 10^5$ m/(As) is the gyromagnetic ratio. The terms on the right-hand-side represent the different torques on the local magnetization.

The first term is the precessional torque,

$$\tau_{precesion} = -\gamma_0 \vec{m} \times \vec{H}_{eff}, \qquad (2)$$

which describes the precession of the magnetization around the local effective field $\vec{H}_{eff}$. This effective field includes exchange, demagnetizing, anisotropy and external field contributions. In the present analysis, a high perpendicular magnetic anisotropy (PMA) strip is considered with the easy axis pointing along the z-axis.

The second term is the damping torque,

$$\vec{\tau}_{damping} = +\alpha\, \vec{m} \times \frac{\partial \vec{m}}{\partial t}, \tag{3}$$

where $\alpha$ is the dimensionless Gilbert damping parameter. This torque describes relaxation of the local magnetization $\vec{m}$ toward its equilibrium, aligned parallel to the local effective field $\vec{H}_{eff}$.

The third term is the spin-transfer torque (STT) that includes both adiabatic and non-adiabatic contributions:

$$\vec{\tau}_{STT} = b_J (\vec{u}_x \cdot \nabla)\vec{m} - \beta\, b_J\, \vec{m} \times (\vec{u}_x \cdot \nabla)\vec{m} \tag{4}$$

where the applied current is $\vec{j}_a = j_a \vec{u}_x$. <u>Note that "current" $j_a$ here denotes conventional current (flow of positive charge carrier), which is in the opposite direction with respect to the *electron flow* $j_e$ in the Letter.</u> The coefficient $b_J$ is given by

$$b_J = \frac{\mu_B P}{e M_s} j_a \tag{5}$$

and $\beta$ is the dimensionless non-adiabatic parameter. $\mu_B = 9.274 \times 10^{-24}$ J/T is the Bohr magneton and $e = -1.6 \times 10^{-19}$ C is the electron charge.

Finally, the last term is the Slonczewski-like torque (SLT) due to the Spin-Hall effect (SHE), which according to Refs. 22 and 23, is given by

$$\vec{\tau}_{SLT}^{SHE} = -\gamma_0 \vec{m} \times \left( \frac{\hbar \theta_{SH} j_a}{2\mu_0 e M_s L_z} \vec{m} \times \vec{u}_y \right), \tag{6}$$

where $\theta_{SH}$ is the spin-Hall angle, and $L_z$ is the thickness of the ferromagnetic sample, and $\hbar = 1.054 \times 10^{-34}$ Js is the Planck's constant. This SHE torque (Eq. 6) can be re-written in a similar form to the standard precession torque (Eq. 2) as

$$\vec{\tau}_{SLT}^{SHE} = -\gamma_0 \vec{m} \times \vec{H}_{SHE} \tag{7}$$

where $\vec{H}_{SHE}$ is the spin Hall effective field, which is the term between brackets in (Eq. 6), and therefore can be written as:

$$\vec{H}_{SHE} = \frac{\hbar \theta_{SH} j_a}{2\mu_0 e M_s L_z} \vec{m} \times \vec{u}_y \tag{8}$$

The direction of this effective field depends on the internal magnetization direction of the Néel DW (Right or Left) and on the magnetization directions of the domains on left and right sides of the DW. This effective field then determines the direction of DW motion (DWM). See discussions below.

## X. Direction of domain wall motion (DWM)

In a thin ferromagnetic strip with high PMA along the $z$-axis, there are two possible senses of magnetization across a DW: (a) Up-Down and (b) Down-Up. These are depicted in the following Fig. S8, along with the coordinate system.

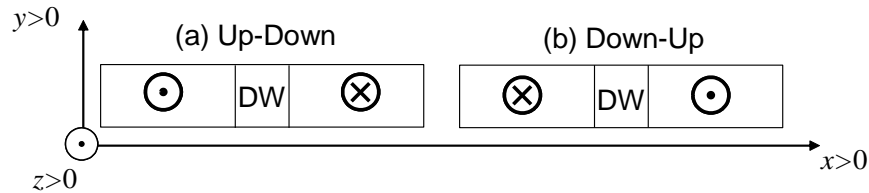

**Figure S8** | DW configurations depending on the magnetization on either side of the DW: Up-Down (**a**) and Down-Up (**b**).

We first consider the simplest case of DW motion driven solely by an out-of-plane (*z*) magnetic field

$$\vec{H}_{ext} = H_{ext}\vec{u}_z, \qquad (9)$$

where $H_{ext}$ can be positive ($\vec{H}_{ext} = +H_{ext}\vec{u}_z$, with $H_{ext}>0$, so pointing along the Up domain) or negative ($\vec{H}_{ext} = -H_{ext}\vec{u}_z$ with $H_{ext}>0$, so pointing along the Down domain). No current is considered in this case so that the STT (Eq. 4) and SLT (Eq. 6) vanish. The combination of the precessional and damping torques drives the magnetization to the direction of $\vec{H}_{ext}$. Therefore, a domain oriented in the direction of $\vec{H}_{ext}$ grows. As summarized in Fig. S9, with $H_{ext}>0$ the Up-Down DW propagates to the right (*x* > 0) and the Down-Up DW propagates to the left (*x* < 0); with $H_{ext}<0$ the Up-Down DW propagates to the left (*x* < 0) and the Down-Up DW propagates to the left (*x* > 0). The magnetization configuration of the DW itself, whether it is Bloch or Néel, does not affect the direction of motion. Thus, two adjacent DWs (one Up-Down and the other necessarily Down-Up) in a magnetic strip propagate in opposite directions under the application of a uniform out-of-plane field (Eq. 9). This feature may not be desirable in many device applications (e.g. DW shift register or "racetrack memory"), because uniform shifting of adjacent digital bits encoded as magnetic domains is required.

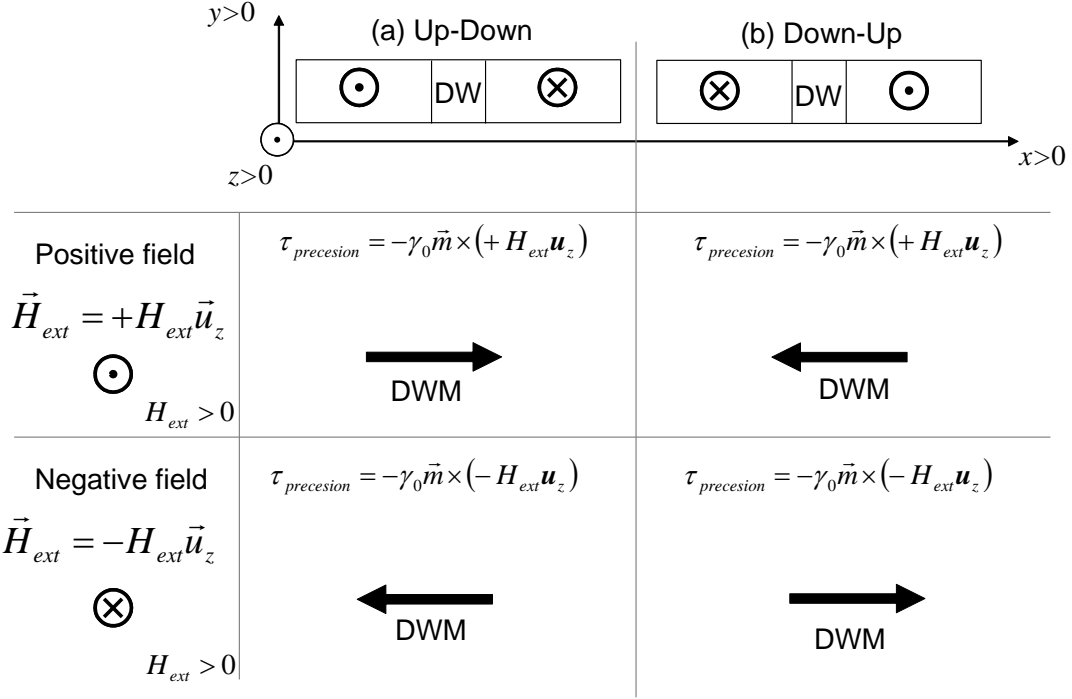

**Figure S9** | Direction of DWM under positive and negative out-of-plane field for the Up-Down (**a**) and Down-Up (**b**) configurations. The direction of field-driven DWM does not depend on the internal magnetization configuration of the DW.

Contrary to the field-driven case, the sense of the current-driven DWM due to the SHE depends on the internal DW magnetization, because the SLT from the SHE depends on the magnetization direction (Eq. 6). The SLT cannot exert a torque on a Bloch DW, whose internal magnetization points along the transverse (*y*) axis, i.e. $\vec{m} \times \vec{u}_y = 0$ in Eq. 6. We therefore focus on the SLT acting on Néel DWs, whose internal magnetization is oriented along the longitudinal (*x*) axis and four possible configurations, as illustrated in Fig. S10.

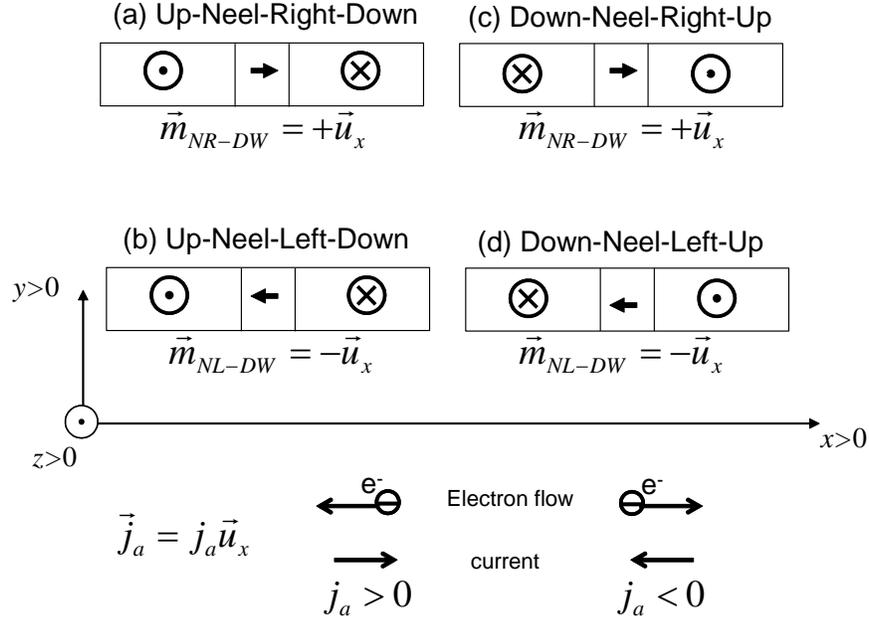

**Figure S10** | DW configurations: Up-Down domains with Neel-Right DW [U-NR-D] (**a**), Up-Down domains with Neel-Left [U-NL-D] (**b**), Down-Up domains with Neel-Right [D-NR-U] (**c**), and Down-Up domains with Neel-Left [D-NL-U] (**d**). The sign criteria for the current and the electron flow are shown at the bottom.

The DWM direction in the presence of the SHE is determined from the direction of the effective field $\vec{H}_{SHE}$ acting on the DW magnetization. In particular, $\vec{H}_{SHE}$ acts in the same way as an applied out-of-plane field only on a Néel DW. (Note that the direction of $\vec{H}_{SHE}$ depends on the magnetization direction $\vec{m}$ as shown in Eq. 8). We show the direction of DWM for each of the possible Néel DW configurations with a positive spin Hall angle ($\theta_{SH} > 0$) and positive conventional current ($j_a > 0$). Note also that the electron charge is a negative quantity here, $e < 0$.

Case 3(a): Up-Neel-Right-Down: $\vec{m}_{NR-DW} = +\vec{u}_x$, so according to (Eq. 8)

$$\vec{H}_{SHE}^{U-NR-D} \approx \frac{\hbar \theta_{SH} j_a}{2\mu_0 e M_s L_z} \vec{m}_{NR-DW} \times \vec{u}_y = \frac{\hbar \theta_{SH} j_a}{2\mu_0 e M_s L_z} \vec{u}_z = -\frac{\hbar \theta_{SH} j_a}{2\mu_0 |e| M_s L_z} \vec{u}_z < 0.$$

That is, $\vec{H}_{SHE}$ acts like an out-of-plane external magnetic field along the negative *z*-axis, favoring growth of the Down domain, which in this case on the right side of the DW. Therefore, the DW moves to the left, against conventional current, in the direction of electron flow.

Case 3(b): Up-Neel-Left-Down: $\vec{m}_{NL-DW} = +\vec{u}_x$, so according to (Eq. 8)

$$\vec{H}_{SHE}^{U-NL-D} \approx \frac{\hbar \theta_{SH} j_a}{2\mu_0 e M_s L_z} \vec{m}_{NL-DW} \times \vec{u}_y = -\frac{\hbar \theta_{SH} j_a}{2\mu_0 e M_s L_z} \vec{u}_z = \frac{\hbar \theta_{SH} j_a}{2\mu_0 |e| M_s L_z} \vec{u}_z > 0.$$

That is, $\vec{H}_{SHE}$ acts like an out-of-plane external magnetic field along the positive *z*-axis, favoring growth of the UP domain, which in this case on the left side of the DW. Therefore, the DW moves to the right, along conventional current, against the direction of electron flow.

Case 3(c): Down-Neel-Right-Up: $\vec{m}_{NR-DW} = +\vec{u}_x$, so according to (Eq. 8)

$$\vec{H}_{SHE}^{D-NR-U} \approx \frac{\hbar \theta_{SH} j_a}{2\mu_0 e M_s L_z} \vec{m}_{NR-DW} \times \vec{u}_y = \frac{\hbar \theta_{SH} j_a}{2\mu_0 e M_s L_z} \vec{u}_z = -\frac{\hbar \theta_{SH} j_a}{2\mu_0 |e| M_s L_z} \vec{u}_z < 0.$$

That is, $\vec{H}_{SHE}$ acts like an out-of-plane external magnetic field along the negative *z*-axis, favoring growth of the Down domain, which in this case on the left side of the DW. Therefore, the DW moves to the right, along conventional current, against the direction of electron flow.

Case 3(d): Down-Neel-Left-Up: $\vec{m}_{NL-DW} = +\vec{u}_x$, so according to (Eq. 8)

$$\vec{H}_{SHE}^{D-NL-U} \approx \frac{\hbar \theta_{SH} j_a}{2\mu_0 e M_s L_z} \vec{m}_{NL-DW} \times \vec{u}_y = -\frac{\hbar \theta_{SH} j_a}{2\mu_0 e M_s L_z} \vec{u}_z = \frac{\hbar \theta_{SH} j_a}{2\mu_0 |e| M_s L_z} \vec{u}_z > 0.$$

That is, $\vec{H}_{SHE}$ acts like an out-of-plane external magnetic field along the positive *z*-axis, favoring growth of the UP domain, which in this case on the right side of the DW. Therefore, the DW moves to the left, against conventional current, in the direction of electron flow.

The above predictions are summarized schematically in Fig. S11.

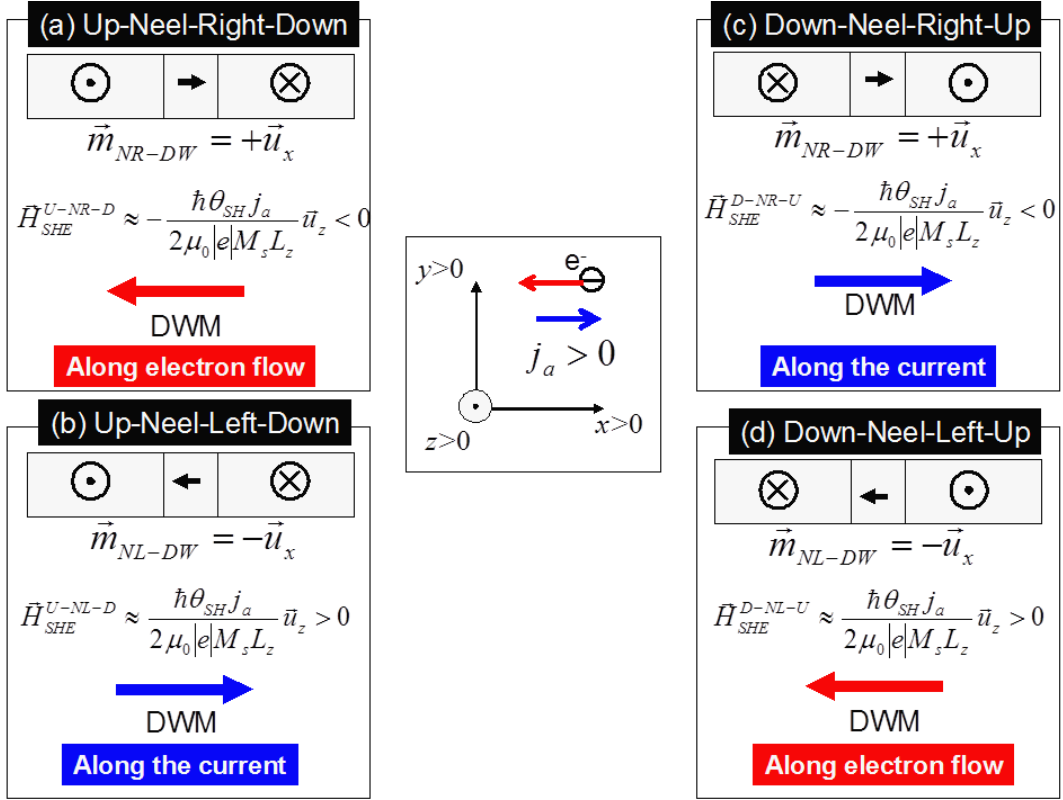

**Figure S11** | Direction of DW motion (DWM) under a positive current ($j_a$>0, in the +x direction) for each of the DW configurations shown in Fig. S10.

The chirality (or the "handedness") of the Néel DW is defined according to the convention illustrated in Fig. S12 and Ref. 21. The Up-Néel-Right-Down (Fig. S11a) and Down-Néel-Left-Up (Fig. S11d) DWs are "Right-Handed," whereas the Up-Néel-Left-Down (Fig. S11b) and Down-Néel-Right-Up (Fig. S11c) DWs are "Left-Handed." With $\theta_{SH} > 0$, the Right-Handed DWs move along electron flow, whereas the Left-Handed DWs move along conventional current. In other words, both Up-Down and Down-Up Néel DWs move in the same direction if they have the same chirality, which is fixed by the DMI. This is a key conclusion of our study as illustrated in Fig. 4 of the Letter. We conclude that the Néel DWs in Pt/CoFe/MgO are Left-Handed, because $\theta_{SH} > 0$ and the direction of DWM is along conventional current. The DWs in Ta/CoFe/MgO must also be Left-Handed, because both $\theta_{SH}$ and DWM direction are opposite with respect to those of Pt/CoFe/MgO.

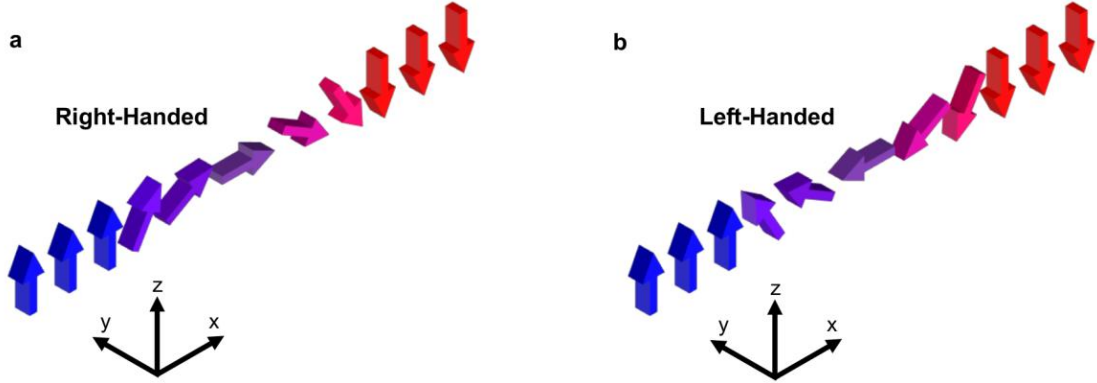

**Figure S12** | Two chiralities of Néel DWs.

## XI. One-dimensional model of domain wall motion

We calculate the DW velocity under different conditions using the one-dimensional model (1DM). The 1DM describes the DW dynamics in terms of two collective coordinates: the DW position $X(t)$ along the strip axis (*x*-axis), and the DW angle $\Phi(t)$, which, as in standard spherical coordinates, here is defined as the in-plane (*xy*) angle with respect to the positive *x*-axis. Therefore, a Néel-Right (NR) DW has its internal magnetization aligned along the positive *x*-axis ($\Phi_{NR} = 0$), and a Néel-Left (NL) is along the negative *x*-axis ($\Phi_{NL} = \pi$), as shown in Figs. S10 and S11. The internal magnetization of a Bloch DW can be either $\Phi_{BU} = \frac{\pi}{2}$ (Bloch-Up) or $\Phi_{BD} = \frac{3\pi}{2}$ (Bloch-Down).

The current $\vec{j}_a = j_a \vec{u}_x$ is injected along the strip/wire axis (*x*-axis), and it is positive ($j_a >$ 0) along the positive *x*-axis ($x > 0$, from left to right). Note that the direction of this (conventional) current $j_a$ is in the opposite direction with respect to the *electron flow* $j_e$ in the Letter. Therefore, the direction of the electron flow for a positive current ($j_a > 0$) is along the negative x-axis ($x < 0$, from right to left). The electric charge $e$ is $-1.6 \times 10^{-19}$ C.

The two coupled differential equations of the 1DM are explained in the Methods section. In addition, the influence of pinning and thermal effects can be taken into account in the 1DM by including the pinning field $H_p(X)$, and the thermal field $H_{th}(t)$ contributions in the out-of-plane field $H_z$, so that $H_z$ is replaced by $H_z + H_p(X) + H_{th}(t)$. The spatially varying pinning field $H_p(X)$ is given by

$$H_p(X) = -\frac{1}{2\mu_0 M_s L_y L_s} \frac{\partial V_{pin}(X)}{\partial X}$$

where $V_{pin}(X)$ is the local pining potential, which is assumed to be described by a periodic function $V_{pin}(X) = V_0 \sin^2(\pi X/p)$, with $V_0$ representing the energy barrier between adjacent minima of the pinning profile and $p$ the spatial periodicity. The values $V_0 = 1.65 \times 10^{-20}$ J and $p = 30$ nm were selected to reproduce the deterministic full micromagnetically computed depinning field for the strip with a characteristic edge roughness size of 3 nm (Ref. 24). The thermal field is given by

$$H_{th}(t) = \eta(t)\sqrt{\frac{2\alpha K_B T}{\gamma_0 \mu_0 M_s L_y L_z \Delta dt}},$$

where $K_B$ is the Boltzmann constant, $dt$ is the temporal step, and $\eta(t)$ is a Gaussian-distributed 1D stochastic process with cero mean value ($<\eta(t)>=0$) and uncorrelated in time ($<\eta(t)\ \eta(t')>=\delta(t-t')$). The coupled 1DM equations were numerically solved by means of a 4$^{th}$ order Runge-Kutta algorithm with $dt = 1$ ps. The deterministic ($T = 0$) results for the DW velocity were computed by considering a temporal window of 100 ns. At room temperature ($T =$ 300 K), the results were averaged over 10 different stochastic realizations for each driving current.

## XII. 1DM results for the DW velocity as function of the current.

Figure S13 shows the DW velocity ($v$) as a function of $j_a$ in several different cases calculated for an Up-Down DW: (a) DWM under STT only: $P = 0.5$, $\xi = 0.5$, $\theta_{SH} = 0$, $D = 0$; (b) DWM under SHE only: $P = 0$, $\theta_{SH} = 0.1$, $D = 0$; and (c) DWM under DMI only: $P = 0$, $\theta_{SH} = 0$, $D = -0.5 \text{mJ/m}^2$. In these calculations, no external field is applied. As it is well-known, if the STT is the only driving force (Fig. S13(a)), the resulting DW motion (DWM) is along electron flow (against conventional current), i.e. $v < 0$ for $j_a > 0$ and $v > 0$ for $j_a < 0$. The SHE alone cannot sustain DW motion (Fig. S13(b)) because it does not have the correct symmetry to exert a torque on a Bloch DW, whose internal DW magnetization (set by the magnetostatics of the srip) is parallel to the *y*-axis. The DMI alone also does not move a DW in the out-of-plane (*z*) magnetized samples (Fig. S13(c)) because it is equivalent to an effective field on the DW along the *x*-axis.

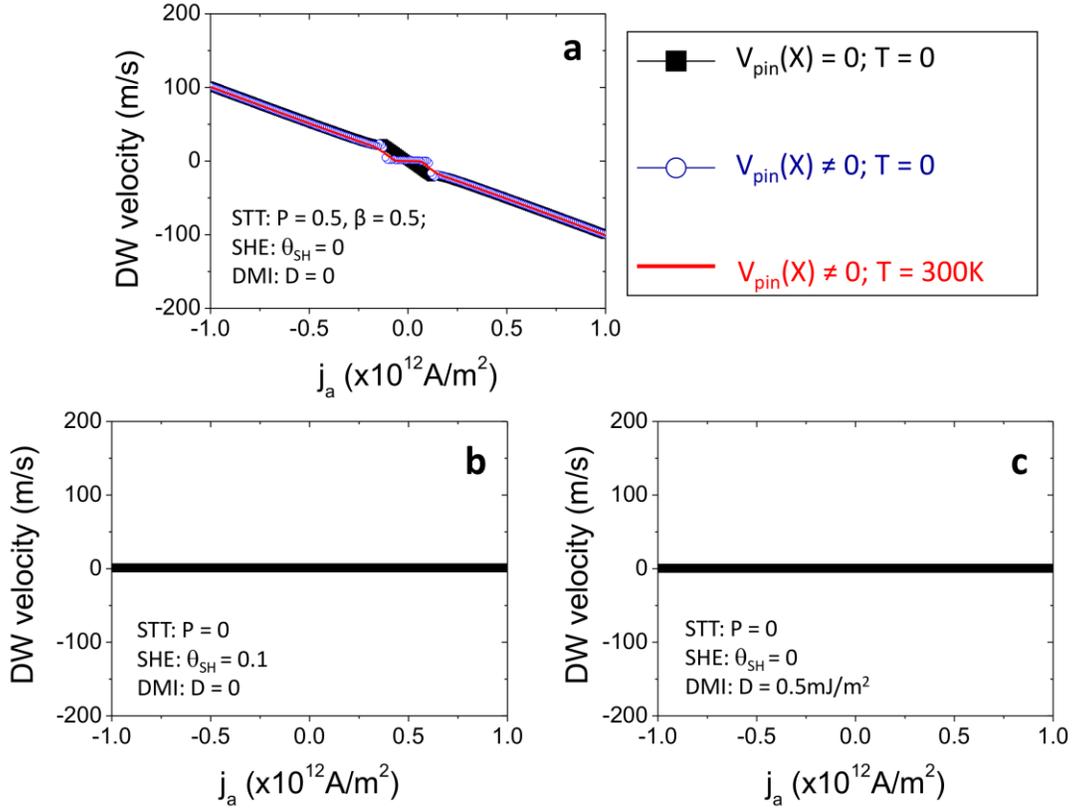

**Figure S13 |** DW velocity as a function of driving conventional current $j_a$ for the cases: (**a**) $P = 0.5$, $\xi = 0.5$, $\theta_{SH} = 0$, $D = 0$; (**b**) $P = 0$, $\theta_{SH} = 0.1$, $D = 0$; and (**c**) $P = 0$, $\theta_{SH} = 0$, $D = -0.5$mJ/m$^2$. Black squares correspond to the perfect sample without pinning ($V_{pin}(X) = 0$) at zero temperature ($T = 0$). Open blue circles are results considering a realistic sample with pinning ($V_{pin}(X) \neq 0$) at $T = 0$, and solid red lines correspond to the realistic sample ($V_{pin}(X) \neq 0$) at $T = 300$K.

Figure S14 shows the DW velocity ($v$) as a function of $j_a$ for different combinations of STT, SHE and DMI torques. The DW dynamics incorporating both the SHE and STT (Fig. S14(a)) are similar to the case with STT (Fig. S13(a)), and DWM is along the electron flow. This is again because the torque from the SHE is null on a Bloch DW. Under the STT and the DMI (Fig. S14(b)), the DW propagates again along electron flow but with higher DW mobility than in former cases, as the DW propagates without precessing and adopts a rigid Néel configuration. The inclusion of the DMI also increases the critical depinning current below which the DW does not propagates at zero temperature (see open blue circles in Fig. S14(b) as compared to Fig. S14(a)).

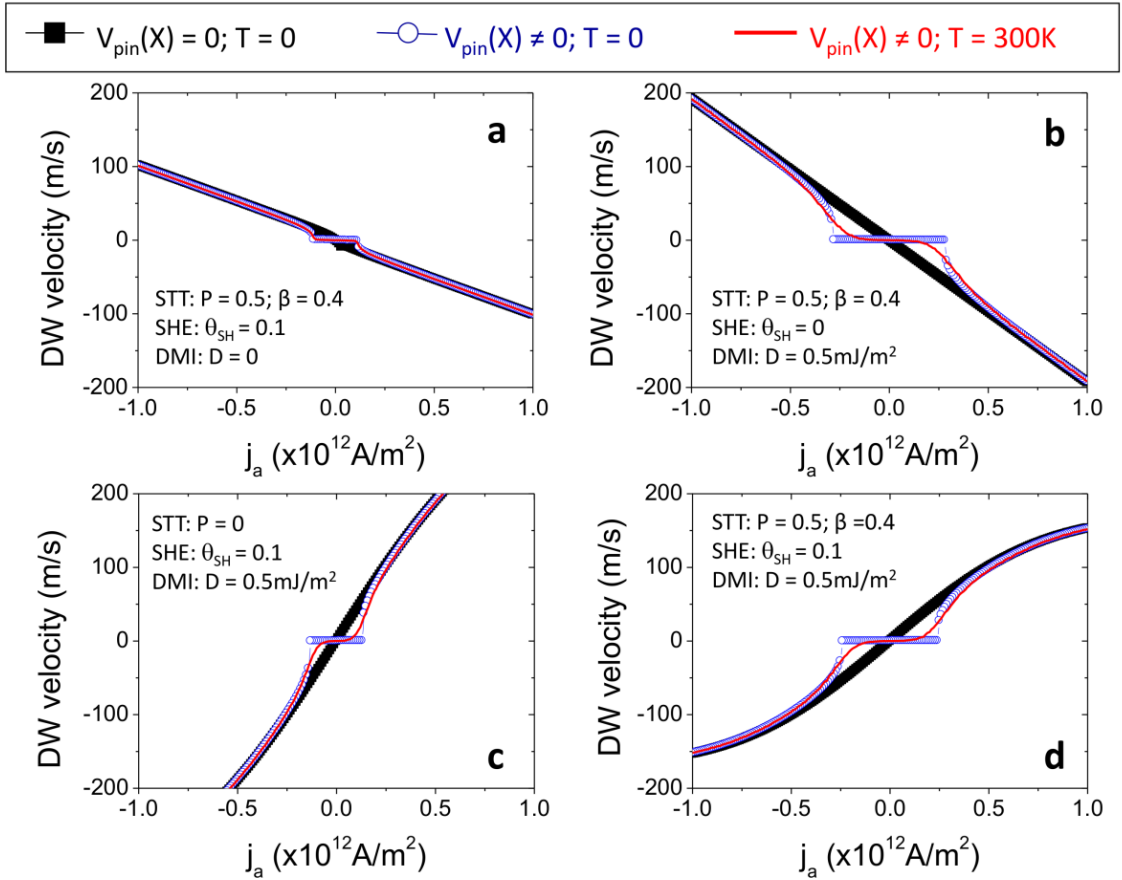

**Figure S14** | DW velocity as a function of driving conventional current $j_a$ for the cases: (**a**) $P = 0.5$, $\xi = 0.5$, $\theta_{SH} = 0.1$, $D = 0$; (**b**) $P = 0.5$, $\xi = 0.5$, $\theta_{SH} = 0$, $D = -0.5$mJ/m$^2$; (**c**) $P = 0$, $\theta_{SH} = 0.1$, $D = 0.5$mJ/m$^2$; and (**d**) $P = 0.5$, $\xi = 0.5$, $\theta_{SH} = 0.1$, $D = -0.5$mJ/m$^2$. Black squares correspond to the perfect sample without pinning ($V_{pin}(X) = 0$) at zero temperature ($T = 0$). Open blue circles are

results considering a realistic sample with pinning ($V_{pin}(X) \neq 0$) at $T = 0$, and solid red lines correspond to the realistic sample ($V_{pin}(X) \neq 0$) at room temperature $T = 300$ K.

When both the SHE and DMI torques are considered (Fig. S14(c)-(d)), the direction of DWM is reversed to along the current (against electron flow), i.e. $v > 0$ for $j_a > 0$. Due to the DMI, in these cases (Fig. S14(c)-(d)) the DW moves rigidly adopting a Néel configuration which is driven by the SHE. The DW mobility is smaller when STT is finite (Fig. S14(d)) as compared to the zero-STT case (Fig. S14(c)), because STT drives the DW into precession and opposes the stabilizing action of the DMI.

## XIII. 1DM results of the current-driven DWM as a function of in-plane field

In Fig. S15, the influence of in-plane fields on the current-driven DW velocity is presented for a perfect sample at zero temperature. Both Longitudinal $H_x \equiv H_L$ and Transversal fields $H_y \equiv H_T$ are evaluated for an Up-Néel-Left-Down DW in the absence of STT ($P = 0$) but taking into account both the SHE ($\theta_{SH} = 0.1$) and the DMI ($D = -0.5$ mJ/m$^2$). At zero longitudinal field ($H_L = 0 = H_x$), the DW propagates along conventional current for both polarities (in the $+x$-direction for $j_a = +0.3 \times 10^{12}$ A/m$^2$ and along the $-x$-direction for $j_a = -0.3 \times 10^{12}$ A/m$^2$. (Note that the case of $j_a = -0.3 \times 10^{12}$ A/m$^2$ is equivalent to the current-driven motion of a "Down-Néel-Right-Up" DW shown in Figs. 4e-h ("down-up DW") of the Letter.)

A negative longitudinal field ($H_L < 0$) supports the DMI ($H_L < 0$ is parallel to the effective DMI field in the Up-Néel-Left-Down DW, $D < 0$), stabilizing the Left-Handed Néel internal DW magnetization. Therefore, $H_L < 0$ does not significantly modify the DW velocity, although a small slope can be observed within a small range around $H_L = 0$ (shown in Fig. 4g of the Letter). When $H_L > 0$, the applied field opposes the DMI with $H_L$ and the effective DMI field antiparallel to each other. A sufficiently large $H_L > 0$ overcomes the DMI and reverses the chirality of the DW, i.e. aligning the internal DW magnetization to the right (Up-Néel-Right-Down). Such reversal of the DW chirality occurs around $H_L = 2000$ Oe in the calculated results (Fig. S14), and this is accompanied by the reversal of the DWM direction.

The influence of the transverse field ($H_T$) on the current-driven DWM is completely different. This in-plane field is orthogonal to the effective DMI field, and it therefore reduces the energy barrier between the Néel and Bloch configurations of the DW. With increasing $|H_T|$, the DW velocity decreases because the DW configuration is pulled away from the ideal Néel configuration that maximizes the Slonczewski-like torque. Under an even larger $|H_T|$ (>10,000 Oe in the calculations, not shown in Fig. S15), the DW adopts a Bloch configuration and the DW velocity approaches zero, as the Bloch DW does not experience a Slonczewski-like torque.

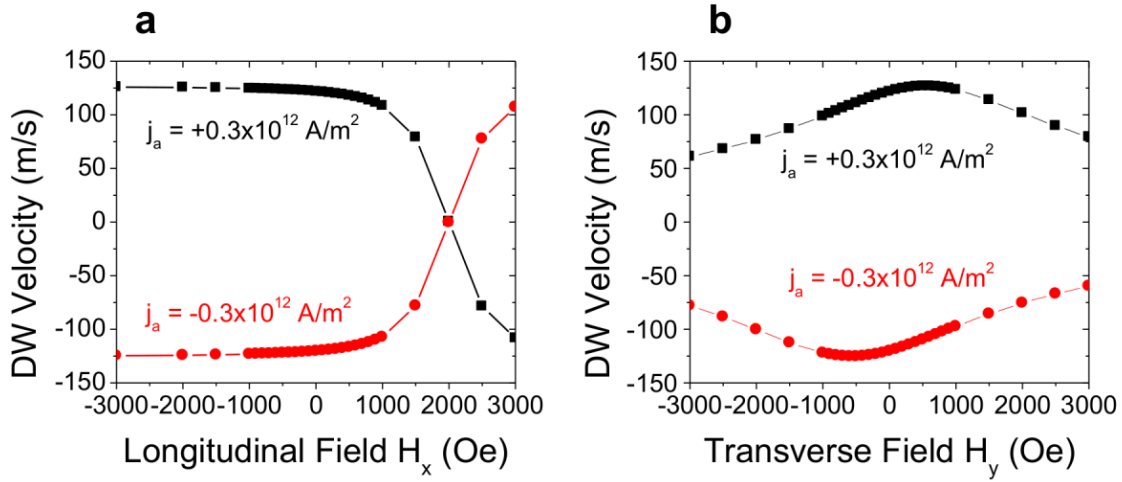

**Figure S15** | Current-driven DW velocity, calculated by the 1DM, as a function of applied in-plane field in the longitudinal direction (**a**) and transverse direction (**b**). A Up-Néel-Left-Down DW at zero temperature and no spatially-varying pinning potential with the following parameters

$P = 0$, $\theta_{SH} = 0.1$, and $D = -0.5$ mJ/m$^2$ is evaluated for $j_a = +0.3\times10^{12}$ A/m$^2$ (black squares) and $j_a = -0.3\times10^{12}$ A/m$^2$ (red circles).

**Supplementary Information References**